\documentclass[11pt]{article}
\usepackage{amssymb,amsmath,amsfonts,graphicx,enumitem, bbm}
\newcommand{\be}{\begin{equation}}
\newcommand{\ee}{\end{equation}}
\newcommand{\bea}{\begin{eqnarray}}
\newcommand{\eea}{\end{eqnarray}}
\newcommand{\beas}{\begin{eqnarray*}}
\newcommand{\eeas}{\end{eqnarray*}}
\def\sqr#1#2{{\vcenter{\vbox{\hrule height.#2pt
\hbox{\vrule width.#2pt height#1pt \kern#1pt
\vrule width.#2pt}\hrule height.#2pt}}}}

\def\identity{{\mathbbmss 1}}

\def\Tr {{\rm Tr}}

\def\del {\partial}

\begin{document}
%%%%%%%%%%%%%%%%%%%%%%%%%%%%%%%%%%%%%%%%%%%%%%%
%\fontfamily{pnb}\fontsize{12pt}{16pt}\selectfont
%\fontfamily{pzc}\fontsize{14pt}{16pt}\selectfont
%\fontfamily{pbk}\fontsize{12pt}{16pt}\selectfont
\fontfamily{cmr}\fontsize{11pt}{16pt}\selectfont
%\fontfamily{phv}\fontshape{ro}\fontsize{11pt}{14pt}\selectfont
%\fontfamily{ptm}\fontseries{m}\fontshape{r}\fontsize{12pt}{16pt}\selectfont
%\fontfamily{pnc}\fontseries{m}\fontshape{r}\fontsize{11pt}{15pt}\selectfont
%\fontfamily{ppl}\fontseries{m}\fontshape{r}\fontsize{11pt}{15pt}\selectfont
%\usefont{T1}{phv}{m}{it}
%%%%%%%%%%%%%%%%%%%%%%%%%%%%%%%%%%%%%%%%%%%%%%%
\def \CMP {{Commun. Math. Phys.}}
\def \PRL {{Phys. Rev. Lett.}}
\def \PL {{Phys. Lett.}}
\def \NPBProc {{Nucl. Phys. B (Proc. Suppl.)}}
\def \NP {{Nucl. Phys.}}
\def \RMP {{Rev. Mod. Phys.}}
\def \JGP {{J. Geom. Phys.}}
\def \CQG {{Class. Quant. Grav.}}
\def \MPL {{Mod. Phys. Lett.}}
\def \IJMP {{ Int. J. Mod. Phys.}}
\def \JHEP {{JHEP}}
\def \PR {{Phys. Rev.}}
\def \JMP {{J. Math. Phys.}}
\def \GRG{{Gen. Rel. Grav.}}
%%%%%%%%%%%%%%%%%%%%%%%%%%%%%%%%%%%%%%%%%%%%%%%
%%%%%%%%%%%%%%%%%%%%%%%%%%%%%%%%%%%%%%%%%%%%%%%
\begin{titlepage}
\null\vspace{-62pt} \pagestyle{empty}
\begin{center}
\rightline{CCNY-HEP-10/1}
\rightline{February 2010}
\vspace{1truein} {\Large\bfseries
Edges and Diffractive Effects in Casimir Energies}\\
\vskip .1in
{\Large\bfseries ~}\\
%%%%%%%%%%%%%%%%%%%%%%%%%%%%%%%%%%%%%%%%%%%%%%%%\vspace{.6in}
{\large DANIEL KABAT$^a$, DIMITRA KARABALI$^a$} and
 {\large V.P. NAIR$^b$}\\
\vskip .2in
{\itshape $^a$Department of Physics and Astronomy\\
Lehman College of the CUNY\\
Bronx, NY 10468}\\
\vskip .1in
{\itshape $^b$Physics Department\\
City College of the CUNY\\
New York, NY 10031}\\
\vskip .1in
\begin{tabular}{r l}
E-mail:&{\fontfamily{cmtt}\fontsize{11pt}{15pt}\selectfont daniel.kabat@lehman.cuny.edu}\\
&{\fontfamily{cmtt}\fontsize{11pt}{15pt}\selectfont dimitra.karabali@lehman.cuny.edu}\\
&{\fontfamily{cmtt}\fontsize{11pt}{15pt}\selectfont vpn@sci.ccny.cuny.edu}
\end{tabular}

\fontfamily{cmr}\fontsize{11pt}{15pt}\selectfont
\vspace{.8in}
%\vspace{1.5in}%\vspace{0.3in}
%%%%%%%%%%%%%%%%%%%%%%%%%%%%%%%%%%%%%%%%%%%%%%%%%%%%%%%%%%%%
\centerline{\large\bf Abstract}
\end{center}
The prototypical Casimir effect arises when a scalar field is confined
between parallel Dirichlet boundaries.  We study corrections to this
when the boundaries themselves have apertures and edges.  We consider several
geometries: a single plate with a slit in it, perpendicular plates
separated by a gap, and two parallel plates, one of which has a 
long slit of large width, related to the case of one plate being semi-infinite.
We develop a general formalism for studying such
problems, based on the wavefunctional for the field in the gap
between the plates.  This formalism leads to a lower dimensional theory
defined on the open regions of the plates or boundaries. The
Casimir energy is then given in terms of the determinant of the nonlocal differential
operator which defines the lower dimensional theory.  
We develop perturbative methods for computing these
determinants.  Our results are in good agreement with known results based on Monte Carlo
simulations. The method is well suited to isolating the diffractive contributions to the Casimir energy.

\end{titlepage}

%%%%%%%%%%%%%%%%%%%%%%%%%%%%%%%%%%%%%%%%%%%%%%%%%%%%%%
\pagestyle{plain} \setcounter{page}{2}
\setcounter{footnote}{0}
\renewcommand\thefootnote{\mbox{\arabic{footnote}}}

%%%%%%%%%%%%%%%%%%%%%%%%%%%%%%%%%%%%%%%%%%%%%%%%%%%%%%%%%%%%%%%%%%%%%%%%%
\section{Introduction}
%%%%%%%%%%%%%%%%%%%%%%%%%%%%%%%%%%%%%%%%%%%%%%%%%%%%%%%%%%%%%%%%%%%%%%%%%

The Casimir effect, as originally conceived, refers to the electromagnetic field
in the presence of two infinite parallel conducting plates.  The plates modify the boundary conditions
on the field in a way which leads to a finite calculable shift in the ground state energy.
Since the original work of Casimir \cite{Casimir:1948dh}, similar effects have been studied
in a variety of different geometries, and a number of different calculational techniques have
been developed.  For a recent review see \cite{Milton:2008st}.

In the present work, we only consider scalar fields for simplicity.
Our goal is to understand what happens when the boundaries themselves
have edges.  For instance, consider an infinite conducting plate with
a hole in it.  How does the size and shape of the hole modify the
ground state energy?  From the point of view of wave mechanics, the
new feature is that the field can undergo diffraction as it passes
through the hole.  So our work could be viewed as the study of
diffractive corrections to Casimir energies. 

An outline of this paper is as follows.  We first develop a general
formalism for studying edge effects, based on writing a
lower-dimensional effective action for the field which lives in the
hole. The total volume in which the field theory is defined is 
considered to be split into separate regions by boundaries, 
some of which have open regions to achieve the required geometry of surfaces.
The lower-dimensional field theory is defined 
on the open regions of boundaries.  
This lower-dimensional action can be obtained by integrating
out the scalar field in the bulk (section \ref{FPIapproach}).  We use
this to study a single plate with a hole in it, and show that the
Casimir energy can be expressed as the determinant of the nonlocal
differential operator which defines the lower dimensional theory (section \ref{single}).  
Our effective action can also be described in terms of the wave functional
of the field, projected
onto the hole, providing a Hamiltonian type of interpretation
where the normal to the surface takes on the role of time
(section \ref{HamiltonianApproach}).  In section
\ref{slit} we specialize to a single plate with a long slit in it.  We
develop perturbative methods for computing the determinants, based on
separating the operator into what could be considered as direct and diffractive
contributions; mathematically these correspond to the``pole'' (quasi-local) and ``cut''
(non-local) contributions in an integral representation.  
In section \ref{perp} these results are used to
obtain the Casimir energy associated with two perpendicular
plates separated by a gap. 
 In section \ref{parallel} we study two parallel plates separated by a gap, obtaining the diffractive contribution to the Casimir energy when one of the plates is semi-infinite.
In both these cases, we compare the values
obtained from our analytical calculation with the numerical 
calculations of the same available in the literature.
The results from the two approaches are in good agreement.
The appendices collect some mathematical results: the
behavior of a field near a single plate with an edge (appendix
\ref{appendix1}) and heat kernels for the Laplacian with periodic
and Dirichlet boundary conditions (appendix \ref{HeatKernels}).

It is important to note that diffraction around solid obstacles does occur in many of the special geometries for which 
exact or close-to-exact results have been found. For the enormous amount of literature
on such cases, we shall refer back to the reviews, except to point out that the multiple scattering
techniques developed by a number of different groups  \cite{MIT1},  \cite{KenKli}, 
\cite{Milton:2008st, Milton2} and the world line techniques of Gies {\it et al}
\cite{gies0} do incorporate diffraction around such solid objects.
Nevertheless, the Casimir energy due to diffraction around edges of openings in boundaries have been calculated
only in a few cases by world line techniques and Monte Carlo simulations \cite{gies}.
The focus of our work is to develop an analytical understanding of such diffractive effects.

%%%%%%%%%%%%%%%%%%%%%%%%%%%%%%%%%%%%%%%%%%%%%%%%%%%%%%%%%%%%%%%
\section{An effective action for edge effects\label{EffectiveAction}}
%%%%%%%%%%%%%%%%%%%%%%%%%%%%%%%%%%%%%%%%%%%%%%%%%%%%%%%%%%%%%%%

\begin{figure}[!b]
\begin{center}
\scalebox{1.2}{\includegraphics{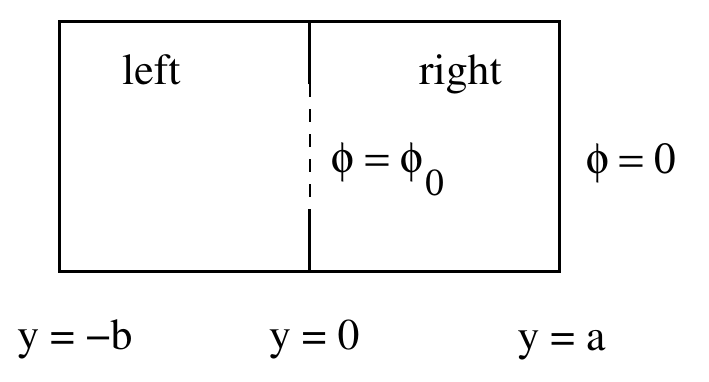}}
\end{center}
\caption{The field propagates in two regions separated by a plate with a hole.
At the location of the hole, indicated by the dotted line, we denote the
fluctuating value of the field by $\phi_0$.  Elsewhere on the boundary, indicated
by solid lines, we impose $\phi = 0$.\label{regions}}
\end{figure}

As a prototype for the sort of problem we will consider, take a free massless scalar field in $d$ Euclidean dimensions.
Imagine it propagates in two regions (``left'' and ``right'')
separated by a plate with a hole in it.  Aside from the hole, we
require that the field vanish everywhere on the boundary, while in the hole, we denote the fluctuating
value of the field by $\phi_0$.  This is illustrated in
Fig.\ \ref{regions}.

The basic idea is to write an effective action $S_0$ for the fluctuations of $\phi_0$.  This effective action can be obtained in two
different ways.  From a path integral point of view it is a lower-dimensional effective action which arises
from integrating out the scalar field in the bulk.  From a Hamiltonian point of view, $S_0$
is related to the wavefunctional of the field projected onto the hole.  We develop the path integral
approach first and return to the Hamiltonian approach in section \ref{HamiltonianApproach}.

%%%%%%%%%%%%%%%%%%%%%%%%%%%%%%%%%%%%%%%%%%%%%%%%%%%%%%%%%%%%%%%%%%
\subsection{Path integral approach\label{FPIapproach}}
%%%%%%%%%%%%%%%%%%%%%%%%%%%%%%%%%%%%%%%%%%%%%%%%%%%%%%%%%%%%%%%%%%

We start with the Euclidean partition function
\be
Z = \int {\cal D} \phi \, e^{- \int d^dx \, {1 \over 2} \partial \phi \cdot \partial \phi}\,
\label{pathint1}
\ee
We fix the value of the field in the hole, $\phi \vert_{\rm hole} = \phi_0$, and subsequently
integrate over $\phi_0$.
\be
Z = \int {\cal D} \phi_0 \, \int_{\raisebox{-5pt}{$\phi \vert_{\rm hole} = \phi_0$}}
\hspace{-15mm} {\cal D} \phi \,\,\,\, e^{- \int d^dx \, {1 \over 2} \partial \phi \cdot \partial \phi}
\label{pathint2}
\ee
To perform the bulk path integral we set
\be
\phi = \phi_{\rm cl} + \delta \phi \,
\label{pathint3}
\ee
Here $\phi_{\rm cl}$ is a solution to the classical equations of
motion $\Box \phi_{\rm cl} = 0$, subject to the boundary conditions
\be
\phi_{\rm cl} = \left\lbrace \begin{array}{ll}
\phi_0 & \hbox{\rm in hole} \\
0 & \hbox{\rm elsewhere on boundary}
\end{array} \right.
\label{pathint4}
\ee
Since $\phi_0$ incorporates the boundary condition, $\delta \phi$ vanishes
on all boundaries including the hole. The action for $\delta\phi$
can then be separated into left and
right regions and the integration over this field can be done.
This leads to
\begin{equation}
\label{Z}
Z = \det{}^{-1/2}(-\Box_L) \det{}^{-1/2}(-\Box_R) \, \int {\cal D} \phi_0 \,
e^{- \int d^dx \, {1 \over 2} \partial \phi_{\rm cl} \cdot \partial \phi_{\rm cl}}
\end{equation}
where $\Box_L,\,\Box_R$ are Laplacians on the left and right
regions.  Given the boundary conditions on $\delta \phi$, they act on
functions that vanish everywhere on the boundary of the left and right
regions (including the hole).

To express $\phi_{\rm cl}$ in terms of $\phi_0$ we introduce $G_L$ and
$G_R$, Green's functions on the left and right.  They obey Dirichlet
boundary conditions: they vanish everywhere on the boundary while in
the bulk they obey $\Box_L G_L(x \vert x') = \delta^d(x-x')$,
$\Box_R G_R(x \vert x') = \delta^d(x-x')$.  In terms of these
Green's functions we have
\begin{equation}
\label{Greens}
\phi_{\rm cl}(x) = \left\lbrace \begin{array}{ll}
\int d^{d-1}x' \, \phi_0(x') \, n \cdot \partial' G_L(x \vert x') & \hbox{\rm on left} \\[8pt]
\int d^{d-1}x' \, \phi_0(x') \, n \cdot \partial' G_R(x \vert x') & \hbox{\rm on right}
\end{array}\right.
\end{equation}
Here $n$ is an outward-pointing unit normal vector.  Integrating by
parts, the classical action in (\ref{Z}) is a surface term which can
be evaluated with the help of (\ref{Greens}).  Putting this all
together, we have
\begin{equation}
\label{Casimir}
Z = \det{}^{-1/2}(-\Box_L) \det{}^{-1/2}(-\Box_R) \, \int {\cal D} \phi_0 \, e^{-S_0}
\end{equation}
where $S_0 = S_L + S_R$ with
\begin{eqnarray}
\label{action1}
S_L & = & \int d^{d-1}x \, \int d^{d-1}x' ~ {1 \over 2} \phi_0(x) \, M_L (x\vert x')~ \phi_0 (x') \\[5pt]
\label{action2}
S_R & = & \int d^{d-1}x \, \int d^{d-1}x' ~ {1 \over 2} \phi_0(x)
\, M_R (x\vert x')~ \phi_0(x')
\end{eqnarray}
and 
\be
M(x\vert x') = n \cdot \partial \, n \cdot \partial' G (x \vert x'),
\label{Mop}
\ee
appropriately for the left and right sides.

The bulk determinants in (\ref{Casimir}) capture the Casimir energy
that would be present if there was no hole.  Corrections to this are
given by a peculiar non-local field theory that lives on the hole
separating the two regions;
the fields $\phi_0$ are nonzero only on the hole. We can write a mode expansion for the fields
$\phi_0$ as
$\phi_0 ({x})
= \sum_\alpha c_\alpha u_\alpha ({ x})$ where $\left\{ u_\alpha ({x})
\right\}$ constitute a complete set of modes for functions which are nonzero
in the hole with the boundary condition that $u_\alpha ({x})\rightarrow 0$
as one approaches the edges of the hole.
The action (\ref{action1}) takes the form
\be
S_L = {1\over 2} \sum_{\alpha,\beta} c_\alpha {\cal O}_{L\alpha\beta} ~c_\beta
\ee
where
\be
{\cal O}_{L\alpha\beta} = \int_{\rm hole} d^{d-1}{x}~d^{d-1}{x'}~ u_\alpha (x) M_L (x\vert x') ~ u_\beta (x')
\ee
with similar expressions for the right side of the partition. 
Because the mode functions $u_\alpha (x)$ vanish outside the hole, this is essentially a
projection of the operator $M(x\vert x')$ to the hole. In other words,
if we define an
operator $P$ which acts on functions $f \in L^2({\mathbb R}^{d-1})$ by
\be
\label{projection}
P f(x) = \left\lbrace
\begin{array}{cl}
f(x) & \hbox{\rm if $x \in {\rm hole}$} \\
0 & \hbox{\rm otherwise}
\end{array}
\right.
\ee
then ${\cal O} = P \, M \, P$.
The functional integration
now leads to
\be
Z = \det{}^{-1/2}(-\Box_L) \det{}^{-1/2}(-\Box_R) \,
\det{}^{-1/2} ( {\cal O}_L + {\cal O}_R)
\label{Casimir2}
\ee

The explicit form of the operator $M(x\vert x')$, and its projected version ${\cal O}$, will, in general,
depend on the arrangement of plates and holes and boundaries. 
We can clarify the nature of this operator by constructing the Green's function with Dirichlet boundary conditions.
For this, consider the right side of the box shown in Fig.\ \ref{regions}.
We split the
coordinates $x = ({\bf x},y)$ into $d-1$ coordinates ${\bf x}$ along
the plate and a single transverse coordinate $y$.  The plate is taken
to be at $y = 0$. (Thus $y$ is along the horizontal axis in the figure.)
The right side of the box has length $a$ along the $y$-direction, while we have
lengths $L_1, L_3, $ etc., along the other directions.
The modes along the $y$-direction are
\be
\psi_n (y) = \sqrt{2\over a} \, ~\sin \left( {n \pi y \over a} \right)
\ee
for $n = 1, 2,$ etc. Similarly modes for the other directions take the form
\be
\psi_{m_i}(x_i) = \sqrt{2\over L_i} \, ~\sin \left( {m_i \pi x_i \over L_i} \right)
\ee
with $m_i = 1,  2$, etc.
The Green's function for the left side can then be written as
\be
G(x\vert x')= - \sum_{\bf p} \psi_{\bf m} ({\bf x}) \psi^*_{\bf m}({\bf x}')
~{2\over a} \sum_n {1\over {\bf p}^2 + n^2 \pi^2 /a^2}
\sin\left( {n \pi y\over a}\right)  \sin\left( {n \pi y' \over a}\right)
\label{green1}
\ee
The summation over $n$ can be carried out by complex integration (or other methods) to obtain
\be
G(x\vert x') =  - \sum_{\bf p}  {1\over 2 p} \psi_{\bf m} ({\bf x}) \psi^*_{\bf m}({\bf x}')
\coth (ap)  \left[ \cosh p (y-y') - \cosh p(y+y')\right]
\label{green2}
\ee
where $p =\sqrt {{\bf p} \cdot {\bf p}}$. This immediately leads to 
\bea
M_R (x\vert x' ) &=& \del_y \del_{y'} G(x\vert x')\nonumber\\
&=& \sum_{\bf p}  \psi_{\bf m} ({\bf x}) \psi^*_{\bf m}({\bf x}')
~\left(p \coth (ap)\right) 
\label{green3}
\eea
Similarly,
\be
M_L (x\vert x') = \sum_{\bf p}  \psi_{\bf m} ({\bf x}) \psi^*_{\bf m}({\bf x}')
~\left(p \coth (b p)\right) 
\label{green4}
\ee
where $b$ is the length (along $y$) of the left side of the box. 
For the parallel plate geometry, we are interested in the limit when
$b \rightarrow \infty$ and $L_i \rightarrow \infty$.
The other cases we shall consider in this paper will also be special cases of the formulae
(\ref{green3}) and (\ref{green4}).
%%%%%%%%%%%%%%%%%%%%%%%%%%%%%%%%%%%%%%%%%%%%%%%%%%%%%%%%%%%%%%%%%
\subsection{Single plate with a hole\label{single}}
%%%%%%%%%%%%%%%%%%%%%%%%%%%%%%%%%%%%%%%%%%%%%%%%%%%%%%%%%%%%%%%%%

We can now go on to the projected version ${\cal O}$ of the operator
$M(x\vert x')$. For this, we will first consider the example of a single plate with a hole in it.
In this case, we are interested in $a\rightarrow \infty$ and $b \rightarrow \infty$.
Further, since $L_i \rightarrow \infty$, we can approximate the sum over
${\bf p}$ by integration,  to obtain
\be
M(x\vert x') =  \int {d^{d-1} p \over (2\pi )^{d-1} }~ e^{i {\bf p}\cdot ({\bf x} - {\bf x}') }
~\sqrt {{\bf p} \cdot {\bf p}}
\ee
This result may also be obtained in a simpler way, without the full mode expansion, by noting that the 
standard Euclidean Green's function is
\[
G({\bf x},y \vert {\bf x}',y') = - {1 \over (d-2) {\rm vol}(S^{d-1}) \Big(\vert {\bf x} - {\bf x}' \vert^2 + (y - y')^2\Big)^{(d-2)/2}}
\]
The Green's function appropriate to the plate geometry (Dirichlet
boundary conditions at $y=0$) can be constructed with the help of an
image charge.
\[
G_D({\bf x},y \vert {\bf x}',y') = G({\bf x},y \vert {\bf x}',y') - G({\bf x},y \vert {\bf x}',-y')
\]
The quantity we need is
\be
\label{singular}
\partial_y \partial_{y'} G_D \vert_{y = y' = 0} = - {2 \over {\rm vol}(S^{d-1})} \,
{\vert {\bf x} - {\bf x}' \vert^2 - (d-1) y^2  \over \left(\vert {\bf x} - {\bf x}' \vert^2 + y^2\right)^{(d+2)/2}}
\ee
where we have kept $y \rightarrow 0^+$ as a regulator.  This is a
quite singular-looking distribution: for ${\bf x} \not= {\bf x}'$ it
approaches $1/\vert {\bf x} - {\bf x}' \vert^d$, while at ${\bf x} =
{\bf x}'$ it diverges as $-1/y^d$.

To interpret this expression we return to the bulk equations of
motion $(\nabla_{\bf x}^2 + \partial_y^2) \phi = 0$. A complete set
of solutions is
\[
\phi_{\rm cl}({\bf x},y) = e^{i {\bf k} \cdot {\bf x}} e^{-ky} \quad \hbox{\rm for $y > 0$}
\]
Notice that $
n \cdot \partial \, \phi_{\rm cl} = - \partial_y \phi_{\rm cl} = k \phi_{\rm cl}$,
so that we can identify
\[
n \cdot \partial \, \phi_{\rm cl} \vert_{y = 0} = \sqrt{-\nabla_{\bf x}^2} \, \phi_{\rm cl} \vert_{y = 0}\,
\]
Denoting the value of the field at $y = 0$ by $\phi_0$, and acting on (\ref{Greens}) with $n \cdot \partial \vert_{y=0}$, this implies that
\be
\label{surprise}
\int d^{d-1}{\bf x}' \, \phi_0({\bf x}') \partial_y \partial_{y'} G_D({\bf x},y \vert {\bf x}',y') \vert_{y = y' = 0} = \sqrt{-\nabla^2}~ \phi_0({\bf x})\,
\ee
In other words, the distribution (\ref{singular}) is the square root of the
Laplacian.  Then the actions (\ref{action1}), (\ref{action2}) are
\be
\label{HoleAction}
S_L = S_R = \int d^{d-1}{\bf x} ~{1 \over 2} \phi_0 \sqrt{-\nabla^2} ~\phi_0
\ee
and the partition function (\ref{Casimir}) is
\be
\label{Casimir3}
Z = \det{}^{-1/2}(-\Box_L) \det{}^{-1/2}(-\Box_R) \, \int {\cal D} \phi_0 \,
\exp \left\lbrace - \int d^{d-1}{\bf x} \, \phi_0 \sqrt{-\nabla^2} ~\phi_0 \right\rbrace
\ee

In this expression $\sqrt{-\nabla^2}$ refers to the
square root of the Laplacian on ${\mathbb R}^{d-1}$, since that is what the arguments leading to (\ref{surprise})
really establish.\footnote{We are grateful to Alexios Polychronakos for discussions on this point.}
But the path integral in (\ref{Casimir3}) is over fields which vanish outside the hole.
Denoting this qualification by the projection operator, the
path integral can be evaluated to give
\[
Z = \det{}^{-1/2}(-\Box_L) \det{}^{-1/2}(-\Box_R) \det{}^{-1/2}\big(P \sqrt{-\nabla^2} P\big)\,
\]
The two bulk determinants can be absorbed by renormalizing the bulk
cosmological constant and the plate tension,\footnote{The stress
tensor associated with an infinite Dirichlet plate is discussed in
Birrell and Davies \cite{Birrell:1982ix}, section 4.3.} so the dependence on the size and
shape of the hole is captured by
\be
\label{Casimir4}
Z = \det{}^{-1/2} \big(P \sqrt{-\nabla^2} P\big)\,
\ee
%%%%%%%%%%%%%%%%%%%%%%%%%%%%%%%%%%%%%%%%%%%%%%%%%%%%%%%%%%%%%%%%%
\subsection{Hamiltonian interpretation\label{HamiltonianApproach}}
%%%%%%%%%%%%%%%%%%%%%%%%%%%%%%%%%%%%%%%%%%%%%%%%%%%%%%%%%%%%%%%%%

For an infinite plate with a hole our expression for the Casimir
energy has a simple Hamiltonian interpretation.  Let us regard $y$ as a
Euclidean time coordinate.  In an unbounded space the vacuum
wavefunctional for the field is given by a Euclidean path integral
over the region $y > 0$.
\[
\Psi_0[\phi_0] = \int_{\raisebox{-5pt}{$\phi({\bf x},y = 0) = \phi_0({\bf x})$}}
\hspace{-25mm} {\cal D} \phi \,\,\,\, e^{- \int d^{d-1}{\bf x} \, \int_0^\infty dy \,
{1 \over 2} \partial \phi \cdot \partial \phi}
\]
Note that $\phi_0$ is defined over the entire ${\bf x}$ plane.
Following the logic in section \ref{FPIapproach}, this means the vacuum wavefunctional
is
\[
\Psi_0[\phi_0] = \det{}^{-1/2}(-\Box_L) \, e^{-S_L[\phi_0]}
\]
which, given (\ref{HoleAction}), can be put in the familiar form
\cite{Jackiw:1988sf,Hatfield:1992rz}
\[
\Psi_0[\phi_0] = \det{}^{-1/2}(-\Box_L) \, e^{- \int d^{d-1}{\bf x} \, {1 \over 2} \phi_0 \sqrt{-\nabla^2} \phi_0}\,.
\]
So our expression (\ref{Casimir}) for the partition function is really
\[
Z = \int_{\raisebox{-5pt}{$\phi_0 = 0$ outside hole}}
\hspace{-25mm} {\cal D} \phi_0 \,\,\,\, \Psi_0^*[\phi_0] \, \Psi_0[\phi_0]
\]
That is, to obtain the partition function we
\begin{enumerate}[itemsep=-.03in]
\item
start with the vacuum state in the far past, at $y = - \infty$
\item
evolve forward in time to $y = 0$
\item
impose a Dirichlet condition by only considering fields which vanish in the region outside the hole
\item
take the overlap with the vacuum state in the far future, evolved backwards in time to $y = 0$
\end{enumerate}
For a single plate with a hole our effective action is related to the vacuum wavefunctional by
\[
e^{-S_0[\phi_0]} = \Psi_0^*[\phi_0] \, \Psi_0[\phi_0]\,.
\]
A similar result holds in general, although with more complicated
plate geometries one no longer has the standard vacuum wavefunctionals
on the left and right.

%%%%%%%%%%%%%%%%%%%%%%%%%%%%%%%%%%%%%%%%%%%%%%%%%%%%%%%%%%%%
\section{Plate with a slit\label{slit}}
%%%%%%%%%%%%%%%%%%%%%%%%%%%%%%%%%%%%%%%%%%%%%%%%%%%%%%%%%%%%

\begin{figure}[!t]
\begin{center}
\includegraphics{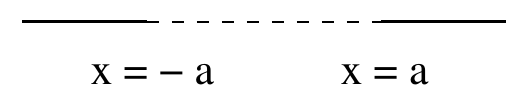}
\end{center}
\caption{In two Euclidean dimensions the slit geometry consists of two whiskers facing each other,
separated by a distance $2a$.\label{SlitGeometry}}
\end{figure}

For a single plate with a hole we have obtained a simple expression for the partition function,
\[
Z = \det{}^{-1/2} \big(P \sqrt{-\nabla^2} P\big)
\]
where $P$ is a projection operator onto the hole and $\sqrt{-\nabla^2}$ is the square root of the
Laplacian on ${\mathbb R}^{d-1}$.  To make further progress we now specialize to the case where the
hole is a long slit of width $2a$.  The geometry is shown in Fig.\ \ref{SlitGeometry}.  We first work
in two dimensions, with a scalar field of mass $\mu$. For a slit in higher dimensions, $\mu$ arises from
Kaluza-Klein momentum along the transverse directions, so we can subsequently integrate over $\mu$ to obtain
results appropriate to the dimension.

In two dimensions the projection operator is
\be
P f(x) = \left\lbrace
\begin{array}{cl}
f(x) & -a < x < a \\
0 & \hbox{\rm otherwise}
\end{array}
\right.
\ee
while $\sqrt{-\nabla^2}$ can be defined by its spectral representation
\[
\sqrt{-\nabla^2} \, e^{i k x} = \vert k \vert \, e^{i k x} \qquad k \in {\mathbb R} \,.
\]
However the operator we need to study is
\[
{\cal O} = P \sqrt{-\nabla^2 + \mu^2}~ P\,.
\]
This is not an easy operator to work with.  In appendix \ref{appendix1} we diagonalize it for a slit
of infinite width.\footnote{Meaning a single plate with an edge, described by
$P f(x) = \left\lbrace
\begin{array}{cl}
f(x) & x > 0 \\
0 & x < 0
\end{array}
\right.$}
But for a slit of finite width we must resort to some sort of approximation scheme.

To do this we note that by construction the field vanishes for $x\geqslant a$ and $x \leqslant -a$.  Moreover ${\cal O}$ respects
a parity symmetry $x \rightarrow -x$.  We therefore expect that we can expand the field in a complete set of odd- and even-parity functions which vanish for $\vert x \vert \geqslant  a$, namely
\bea
\label{OddModeFtns}
&& \psi^{\rm odd}_m = \left\lbrace
\begin{array}{cl}
(-1)^m {1\over \sqrt{a}} \sin \left(m \pi x / a\right) & \hbox{\rm for $-a \leqslant  x \leqslant a$} \\
0 & \hbox{\rm otherwise}
\end{array}\right. \\
\label{EvenModeFtns}
&& \psi^{\rm even}_p = \left\lbrace
\begin{array}{cl}
(-1)^{p + {1\over 2}} {1\over \sqrt{a}} \cos \left(p \pi x / a\right) & \hbox{\rm for $-a \leqslant  x \leqslant a$} \\
0 & \hbox{\rm otherwise}
\end{array}\right.
\eea
The odd modes are labeled by $n,m = 1,2,3,\ldots$ while the even modes carry an index $p,q = {1 \over 2},{3 \over 2},{5 \over 2},\ldots$.  These modes are orthonormal; the factors of $(-1)^m$ and
$(-1)^{p + {1\over 2}}$ are inserted
for later convenience.

These modes are eigenfunctions of the Laplacian with Dirichlet
boundary conditions at $x=a$ and $x=-a$.  We will use them as a basis
in which to diagonalize ${\cal O} = P \sqrt{-\nabla^2 + \mu^2}
~P$.\footnote{One might question whether these modes provide a good
basis in which to diagonalize ${\cal O}$.  This seems justified by the
results of appendix \ref{appendix1}, where we show that the exact
eigenfunctions of ${\cal O}$ indeed go to zero as one approaches the edge of
the slit (in fact they vanish as the square root
of the distance from the edge).}  In this basis it turns out that
${\cal O}$ naturally splits into two pieces: a ``pole'' piece which is
quasi-local and can be diagonalized, and a ``cut'' piece which is
truly non-local.  

As a guide for the reader, in section \ref{PoleCut} we consider the
decomposition of ${\cal O}$ into its pole and cut
contributions.  In section \ref{SlitPert} we set up the perturbation series and
derive integral expressions for all higher terms in this expansion of the partition function.
In section \ref{4dSlit} we integrate over the mass to find the ground state energy for a
plate with a slit in four dimensions.

%%%%%%%%%%%%%%%%%%%%%%%%%%%%%%%%%%%%%%%%%%%%%%%%%%%%%%%%%%%%%%%%%%
\subsection{Pole and cut contributions\label{PoleCut}}
%%%%%%%%%%%%%%%%%%%%%%%%%%%%%%%%%%%%%%%%%%%%%%%%%%%%%%%%%%%%%%%%%%

We begin by considering the odd-parity modes (\ref{OddModeFtns}).  They have a Fourier sine representation
\[
\psi^{\rm odd}_n = {2 \sqrt{a} \over \pi} \int_0^\infty dk \, {n \pi \over k^2 a^2 - n^2 \pi^2} \sin(ka) \sin(kx)\,
\]
Clearly $P \psi^{\rm odd}_n = \psi^{\rm odd}_n$, while for any function of the Laplacian we have
\[
F\big(-\nabla^2\big) \psi^{\rm odd}_n = {2 \sqrt{a} \over \pi} \int_0^\infty dk \, {n \pi F(k^2) \over k^2 a^2 - n^2 \pi^2} \sin(ka) \sin(kx)\,
\]
It follows that the matrix elements in this basis are
\beas
{\cal O}^{\rm odd}_{mn} & = & \langle m \vert P \sqrt{-\nabla^2 + \mu^2}~ P \vert n \rangle \\
& = & {2 a \over \pi} \int_{-\infty}^\infty dk \, \sin^2(ka)\, {m \pi \over k^2 a^2 - m^2 \pi^2} \, \sqrt{k^2+\mu^2}
~{n \pi \over k^2 a^2 - n^2 \pi^2}
\eeas
Despite appearances, there are no singularities on the contour of
integration: the would-be poles at $k = \pm m \pi / a$ and $\pm n \pi
/ a$ cancel against the zeroes of $\sin(ka)$.
\begin{figure}[!t]
\begin{center}
\includegraphics{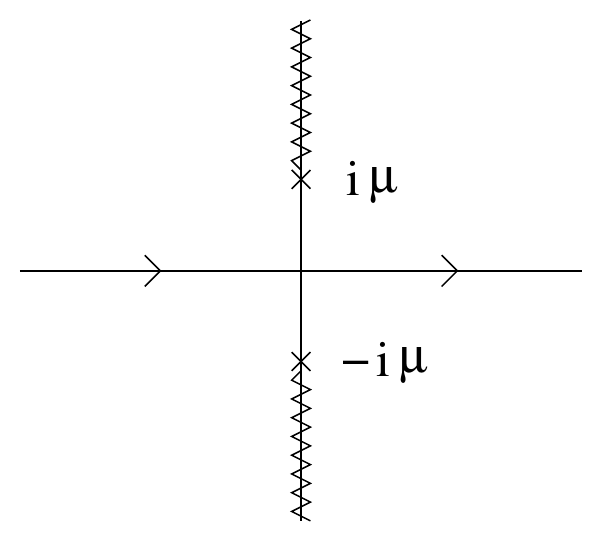}
\end{center}
\caption{The contour for integrating over $k$.}
\end{figure}

Although ${\cal O}$ is not diagonal in this basis, the diagonal matrix elements are numerically
much larger than the
off-diagonal elements.  There is a way of decomposing ${\cal O}$
which makes this manifest.  First deform the integration contour
slightly, moving it just above the real $k$ axis.  Then write
\[
\sin^2(ka) = - {1 \over 4} \left(e^{2ika} + e^{-2ika} - 2\right)\,
\]
For each term in this decomposition the integration contour can be
deformed into the upper or lower half plane.  One picks up a
contribution if the integration contour crosses the poles (now real)
at $k = \pm m \pi / a$ or $\pm n \pi / a$.  One also gets a contribution
when the contours get wrapped around the cuts.
The residues turn out to cancel unless $m = n$, so the pole contribution
to the matrix element is diagonal.  In fact
\be
\label{Opolemn}
{\cal O}^{\rm odd,pole}_{mn} = {1 \over a} \sqrt{n^2 \pi^2 + \mu^2 a^2} \, \delta_{mn}\,
\ee
The cut contribution to the matrix element is not diagonal.  Rather we find
\be
\label{Ocutmn}
{\cal O}^{\rm odd, cut}_{mn} = - {2 \mu^2 a \over \pi} \int_1^\infty dy \,
\sqrt{y^2 - 1} \left(1 - e^{-2\mu ay}\right)
{m \pi \over m^2 \pi^2 + \mu^2 a^2 y^2} \, {n \pi \over n^2 \pi^2 + \mu^2 a^2 y^2}
\ee
Here $y = {\rm Im} \, k / \mu$ is an integration variable along the cut.

Likewise the even-parity modes (\ref{EvenModeFtns}) have a Fourier cosine representation
\[
\psi_p^{\rm even} = {2 \sqrt{a} \over \pi} \int_0^\infty dk \, {p \pi \over k^2 a^2 - p^2 \pi^2} \,
\cos(ka) \cos(kx)
\]
and the matrix elements of ${\cal O}$ in the even-parity sector are
\[
{\cal O}^{\rm even}_{pq} = {2 a \over \pi} \int_{-\infty}^\infty dk \, \cos^2(ka)
{p \pi \over k^2 a^2 - p^2 \pi^2} \sqrt{k^2 + \mu^2} {q \pi \over k^2 a^2 - q^2 \pi^2} \,
\]
Deforming contours as before leads to the decomposition
${\cal O}_{pq} = {\cal O}^{\rm pole}_{pq} + {\cal O}^{\rm cut}_{pq}$ where
\bea
&& {\cal O}^{\rm even,pole}_{pq} = {1 \over a} \sqrt{p^2 \pi^2 + \mu^2 a^2} \, \delta_{pq} \\
\nonumber
&& {\cal O}^{\rm even,cut}_{pq} = - {2 \mu^2 a \over \pi} \int_1^\infty dy \, \sqrt{y^2-1}
\left(1 + e^{-2\mu ay}\right) {p \pi \over p^2 \pi^2 + \mu^2 a^2 y^2} \, {q \pi \over q^2 \pi^2 + \mu^2 a^2 y^2}
\eea
Note the opposite sign in front of the exponential, due to the fact that the even-parity matrix
elements involved $\cos^2 ka$ rather than $\sin^2 ka$.

Combining the even- and odd-parity matrix elements, note that ${\cal O}^{\rm pole}$ can be identified
with the operator $\sqrt{-\nabla_D^2 + \mu^2}$,
where $\nabla^2_D$ denotes the Laplacian with Dirichlet boundary conditions at $x = \pm a$.
So although ${\cal O}^{\rm pole}$ is not a local differential operator, its square is local, and
in this sense we will refer to ${\cal O}^{\rm pole}$ as being quasi-local.  The cut contributions, on the other hand, make a truly nonlocal contribution to the operator ${\cal O}$.

\begin{figure}[!t]
\begin{minipage}{6.5cm}
\begin{center}
\includegraphics{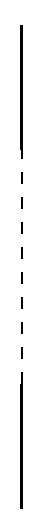}
\end{center}
\end{minipage}
\begin{minipage}{6cm}
\begin{center}
\includegraphics{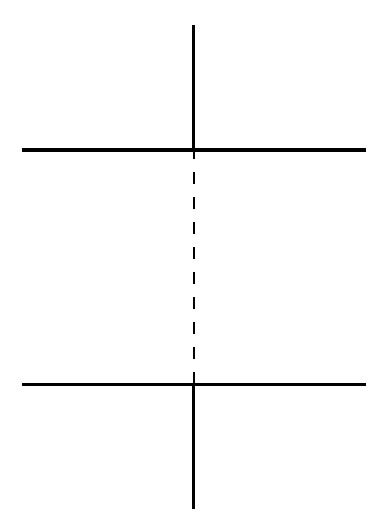}
\end{center}
\end{minipage}
\caption{On the left, the geometry of interest.  On the right, the geometry described
by ${\cal O}^{\rm pole}$, in which additional plates have been added to enforce Dirichlet
boundary conditions at the edges of the hole.\label{DirichletPlates}}
\end{figure}

From the physical point of view the decomposition into pole and cut contributions is natural because
${\cal O}^{\rm pole}$ captures the geometrical optics effects of the hole, in which waves are directly
transmitted from left to right, while
${\cal O}^{\rm cut}$ captures the diffractive effects.  This follows from the observation
made above, that ${\cal O}^{\rm pole}$ is related to an operator with Dirichlet boundary conditions
at the edges of the hole.  Such boundary conditions could be enforced by introducing additional plates
as shown in Fig.\ \ref{DirichletPlates}.  The additional plates prevent any diffraction from taking place,
so diffractive effects are entirely encoded in ${\cal O}^{\rm cut}$.

%%%%%%%%%%%%%%%%%%%%%%%%%%%%%%%%%%%%%%%%%%%%%%%%%%%%%%%%%%%%%%%%%%
\subsection{Perturbation expansion\label{SlitPert}}
%%%%%%%%%%%%%%%%%%%%%%%%%%%%%%%%%%%%%%%%%%%%%%%%%%%%%%%%%%%%%%%%%%

Having decomposed the operator ${\cal O} = {\cal O}^{\rm pole} + {\cal O}^{\rm cut}$ into direct
and diffractive contributions, we wish to find a similar decomposition of the Casimir energy.
This is straightforward.  Expanding in powers of ${\cal O}^{\rm cut}$, the partition function is
\bea
- \log Z & = & {1 \over 2} {\rm Tr} \, \log \left({\cal O}^{\rm pole} + {\cal O}^{\rm cut}\right) \nonumber\\
& = & {1 \over 2} {\rm Tr} \, \log {\cal O}^{\rm pole} + {1 \over 2} {\rm Tr} \, {\cal O}_{\rm pole}^{-1} {\cal O}_{\rm cut}
- {1 \over 4} {\rm Tr} \, {\cal O}_{\rm pole}^{-1} {\cal O}_{\rm cut} {\cal O}_{\rm pole}^{-1} {\cal O}_{\rm cut} + \cdots
\label{pertexp1}
\eea
The zeroth order term in this expansion gives the direct contribution to the energy, while the first and
higher order terms give the diffractive contribution.

Writing things in this way, the diffractive contribution to the energy is organized as a series expansion
in powers of ${\cal O}^{\rm cut}$.  This expansion seems to be
well behaved, even though there is no small parameter in the problem.\footnote{Note that the diffractive contribution to the
energy does not in general have to be small compared to the direct contribution.  Rather what we are claiming is that the
diffractive contribution by itself has a useful series expansion in powers of ${\cal O}^{\rm cut}$.}
We give a speculative reason for this in the conclusions.  But more prosaically the good behavior of the perturbation
series will become evident
from the explicit
calculations we perform in the remainder of this paper, where we work up to 5${}^{th}$ order in ${\cal O}^{\rm cut}$.  For a graphical
preview of the results see Fig.\ \ref{2dslit}.

The lowest-order term in the perturbation series (\ref{pertexp1}), the direct term,  has a  
linear divergence and a subdominant logarithmic divergence, while all higher order terms, corresponding to diffractive contributions, are logarithmically  divergent.
These logarithmic divergences are independent of $a$ and can be eliminated by subtracting the
$a\rightarrow \infty$ limit.
This can be done either from the beginning, before the expansion 
in powers of ${\cal O}^{\rm cut}$, or at the level of each term in the expansion.

These subtractions can be interpreted as renormalizations of parameters corresponding to the plates and slits. Strictly speaking, in addition to the action for the fields, we have an action which describes the plates and slits in the given arrangement. This part of the action is generally of the form
\be
S = \sigma~ {\cal A} ~+~ \alpha~ {\cal L} ~+ \cdots\label{pertexp1.0}
\ee
where ${\cal A}$ is the area of the plate, ${\cal L}$ is the length of the perimeters involved
(for the plate and for any slits or holes in it). The coefficients $\sigma$ and $\alpha$ are the tensions for the plate and the edges of the slits. 
Being the coefficients of the area and perimeter terms, in the language of general relativity,
they are the cosmological constants for the plate and for the  boundaries.
While these are calculable in terms of the material properties of the plates, at the level we are working, with the effects of the plates introduced as merely boundary conditions, they are free parameters.
The partition function and the free energy we calculate are to be thought of as giving corrections to this action (\ref{pertexp1.0}).
The divergent terms we find can be absorbed as renormalizations of the parameters 
$\sigma$, $\alpha$. (In reality, at very short distances, the atomic structure of the plates become important and the divergent terms are rendered finite and calculable in terms of the
interactions at that scale.) When there is a slit or hole in the plate, there is a part of the $\sigma
{\cal A}$ term missing and the renormalization of $\sigma$ appears in a way that depends on the
dimensions of the hole or slit. This is because the term corresponding to the full
area of the  plates (ignoring holes and slits) is already subtracted out as explained at the end of subsection \ref{single}
and, so, the deficit is what is relevant for this part.
The perimeter term should not depend on $a$, but only on the measure of the boundary
(which is just two points for the one-dimensional slit) and can be identified easily by taking
the large $a$ limit.

%%%%%%%%%%%%%%%%%%%%%%%%%%%%%%%%%%%%%%%%%%%%%%%%%%%%%%%%%%%%
\subsubsection{The direct contribution (lowest order)\label{lowestpert}}
%%%%%%%%%%%%%%%%%%%%%%%%%%%%%%%%%%%%%%%%%%%%%%%%%%%%%%%%%%%%

We now proceed to study the various terms in (\ref{pertexp1}).
Combining the even- and odd-parity contributions, at leading order we have
\bea
- \log Z^{(0)} & = & {1 \over 2} \sum_{r = {1 \over 2},1,{3 \over 2},2,\cdots}
\log \sqrt{(r \pi / a)^2 + \mu^2} \nonumber\\
& = & {1 \over 4} \, {\rm Tr} \, \log \left(-\nabla^2_D + \mu^2\right)
\label{pertexp1a}
\eea
where again $\nabla^2_D$ is the Laplacian in the slit with Dirichlet boundary conditions.
This can be computed using the heat kernel methods described in appendix \ref{HeatKernels}.  There is
a linear divergence, proportional to $a$, which renormalizes the cosmological constant in the slit.
There is also a log divergence, independent of $a$, which renormalizes the boundary cosmological
constant (i.e.\ the tension associated with the edges of the slit).  After these divergences are removed
one is left with a finite result which vanishes exponentially as $a \rightarrow \infty$.  In the notation
of appendix \ref{HeatKernels},
\bea
\label{2dlowest}
- \log Z^{(0)}_{\rm renormalized} & = & - {1 \over 4} \int_0^\infty {ds \over s} e^{-s \mu^2}
\left(K_D(s,2a) - {a \over \sqrt{\pi s}} + {1 \over 2}\right) \nonumber\\
& = & - {a \over 2 \sqrt{\pi}} \int_0^\infty {ds \over s^{3/2}} e^{-s \mu^2} \sum_{n = 1}^\infty
e^{- 4 a^2 n^2 / s} \nonumber\\
\label{Z0renormalized}
& = & {1 \over 4} \log \left(1 - e^{-4 \mu a}\right)
\eea

%%%%%%%%%%%%%%%%%%%%%%%%%%%%%%%%%%%%%%%%%%%%%%%%%%%%%%%%%%%%
\subsubsection{The diffractive contributions}
%%%%%%%%%%%%%%%%%%%%%%%%%%%%%%%%%%%%%%%%%%%%%%%%%%%%%%%%%%%%

For the first order term it is useful to separate the contributions from the odd and even parity terms as
\be
-\log Z^{(1)} = - \log Z^{(1)}_{\rm odd} ~- \log Z^{(1)}_{\rm even}
\label{slit-diff1}
\ee
where
\bea
- \log Z^{(1)}_{\rm odd} \!\!\!\!&=&\!\!\!\! - {(\mu a)^2 \over \pi} \int_1^\infty dy \sqrt{y^2-1}~\sum_{r=1,2,\cdots}
{ (1- e^{-2 \mu a y})\,  r^2 \pi^2 \over \sqrt{r^2\pi^2 + \mu^2 a^2}~ ( r^2 \pi^2 + \mu^2 a^2 y^2 )^2}
\label{slit-diff2a}\\
- \log Z^{(1)}_{\rm even} \!\!\!\!&=&\!\!\!\! - {(\mu a)^2 \over \pi} \int_1^\infty dy \sqrt{y^2-1}~\sum_{r={1\over 2},
{3\over 2},\cdots}
{ (1+ e^{-2 \mu a y}) \, r^2 \pi^2 \over \sqrt{r^2\pi^2 + \mu^2 a^2} ( r^2 \pi^2 + \mu^2 a^2 y^2 )^2}
\label{slit-diff2b}
\eea
For ease of presentation of these and higher order results, we define
\bea
T (\mu a, y, z) \!\!\!\!&=&\!\!\!\!  {\mu^2 a^2 \over \pi} \sum_{r=1,2,\cdots}
{  r^2 \pi^2 \over \sqrt{r^2\pi^2 + \mu^2 a^2}~ ( r^2 \pi^2 + \mu^2 a^2 y^2 )~( r^2 \pi^2 + \mu^2 a^2 z^2)}\nonumber\\
\!\!\!\!&=&\!\!\!\! {\mu^2 a^2 \over \pi^2}\sum \int_{-\infty}^\infty\!\! d\lambda 
{  r^2 \pi^2 \over (r^2\pi^2 + \mu^2 a^2 +\lambda^2)~ ( r^2 \pi^2 + \mu^2 a^2 y^2 )~( r^2 \pi^2 + \mu^2 a^2 z^2)}\label{slit-diff3}
\eea
By resolving the integrand into partial fractions, the summation can be done using
\be
\sum_r {1\over r^2 \pi^2 +A^2} = {1\over 2 A^2} \left[ A \coth A - 1\right]
\label{slit-diff4}
\ee
We can then write $T$ as
\bea
T (\mu a, y, z) \!\!\!&=&\!\!\! {1\over \pi^2} \int_0^\infty d\lambda \Biggl[
- {\sqrt{\lambda^2 +1}\coth(\mu a \sqrt{\lambda^2 +1}) \over (\lambda^2 +1 -y^2) (\lambda^2 +1 -z^2)}\nonumber\\
&&\hskip .3in + { y \coth (\mu a y) \over (y^2 -z^2) ( \lambda^2 +1 - y^2)}
+ { z \coth (\mu a z) \over (z^2 -y^2) ( \lambda^2 +1 - z^2)}
\Biggr]\label{slit-diff5}
\eea
The limit of $z\rightarrow y$ is seen to be
\be
T (\mu a, y, y) = {1\over 2\pi^2 y} {d\over dy} \int_0^\infty d\lambda 
\left[ {-\sqrt{\lambda^2 +1}\coth(\mu a \sqrt{\lambda^2 +1}) + y \coth (\mu a y) \over
\lambda^2 +1 -y^2}\right]
\label{slit-diff6}
\ee
The $a\rightarrow \infty$ limit of these expressions is logarithmically divergent.
The renormalized contribution is obtained after subtraction of this divergence as
\be
-\log Z^{(1)}_{\rm odd, ren} = 
- \int_1^\infty dy \sqrt{y^2-1}\, \left[ ( 1- e^{- 2\mu a y}) \,T(\mu a, y, y) -
T(\mu a \rightarrow \infty, y, y)\right]
\label{slit-diff7}
\ee
Similarly, for the even-parity contribution, we define
\bea
S(\mu a, y, z)\! \!\!\!&=&\!\!\! \!{\mu^2 a^2 \over \pi} \sum_{r = {1\over 2}, {3\over 2}, \cdots}
{r^2 \pi^2 \over \sqrt{r^2\pi^2 +\mu^2 a^2}~ (r^2\pi^2 + \mu^2 a^2 y^2) ~(r^2\pi^2
+\mu^2a^2 z^2)}\nonumber\\
\!\!\!\!&=&\!\!\!\! 2 {(2\mu a)^2 \over \pi^2} \sum_{l= {\rm odd}}
\int_{-\infty}^\infty \!\!d\lambda 
{ l^2 \pi^2 \over (l^2\pi^2 + (2\mu a)^2 + \lambda^2) (l^2\pi^2 + (2\mu a)^2 y^2)
(l^2\pi^2 + (2\mu a)^2 z^2)}\nonumber\\
\label{slit-diff8}
\eea
Again, by use of partial fractions and (\ref{slit-diff4}), we can write this as
\bea
S(\mu a, y, z)\!\!\!\!&=&\!\!\!\! {1\over \pi^2} \int_0^\infty d\lambda \Biggl[
- {\sqrt{\lambda^2 +1}\, [ \coth(2 \mu a \sqrt{\lambda^2 +1}) - \text{csch} (2 \mu a \sqrt{\lambda^2 +1}) ]\over (\lambda^2 +1 -y^2) (\lambda^2 +1 -z^2)}\nonumber\\
&&\hskip .3in + { y [\coth (2\mu a y) - \text{csch}(2 \mu a y)]\over (y^2 -z^2) ( \lambda^2 +1 - y^2)}
+ { z [\coth (2\mu a z) - \text{csch}(2 \mu a z)]\over (z^2 -y^2) ( \lambda^2 +1 - z^2)}
\Biggr]\label{slit-diff9}
\eea
\bea
S(\mu a, y, y) \!\!\!\!&=&\!\!\!\! {1\over 2\pi^2 y} {d \over dy}\int_0^\infty d \lambda 
\Biggl[-{ \sqrt{\lambda^2 +1}\, [ \coth(2 \mu a \sqrt{\lambda^2 +1}) - \text{csch} (2 \mu a \sqrt{\lambda^2 +1}) ]\over \lambda^2 +1 -y^2}\nonumber\\
&&\hskip 1in+ { y [\coth (2\mu a y) - \text{csch}(2 \mu a y)] \over \lambda^2 +1 - y^2}
 \Biggr]\label{slit-diff10}
\eea
The renormalized expression for the even-parity contribution is then
\be
-\log Z^{(1)}_{\rm even, ren} = 
- \int_1^\infty dy \sqrt{y^2-1} \left[ (1+ e^{- 2\mu a y}) \,S(\mu a, y, y) -
S(\mu a\rightarrow \infty, y, y)\right] \label{slit-diff11}
\ee

The higher order terms can also be written down easily in terms of 
$T(\mu a, y, z)$ and $S(\mu a, y, z)$ as
\bea
-\log Z^{(n)}_{\rm odd} \!\!\!\!&=&\!\!\!\! 
- {2^{n-1}\over n} \int_1^\infty \prod^n_i dy_i \sqrt{y_i^2-1} \Biggl[
\prod^n_i (1- e^{-2 \mu a y_i}) T(\mu a)\ast T(\mu a)\ast \cdots
\ast T(\mu a) \nonumber\\
&&\hskip 1.7in-  (\mu a \rightarrow \infty)\Biggr]\label{slit-diff12a}\\
-\log Z^{(n)}_{\rm even} \!\!\!\!&=&\!\!\!\! 
- {2^{n-1}\over n} \int_1^\infty \prod^n_i dy_i \sqrt{y_i^2-1} \Biggl[
\prod^n_i (1+ e^{-2 \mu a y_i}) S(\mu a)\ast S(\mu a)\ast \cdots
\ast S(\mu a) \nonumber\\
&&\hskip 1.7in-  (\mu a \rightarrow \infty )\Biggr]\label{slit-diff12b}
\eea
where
\be
T\ast T \ast \cdots \ast T = T(\mu a, y_1, y_2) T(\mu a, y_2, y_3) \cdots
T(\mu a, y_n, y_1),
\label{slit-diff13}
\ee
with a similar expression for the $S$'s. 

\begin{figure}[!t]
\begin{center}
\scalebox{.9}{\includegraphics[width = 10cm]{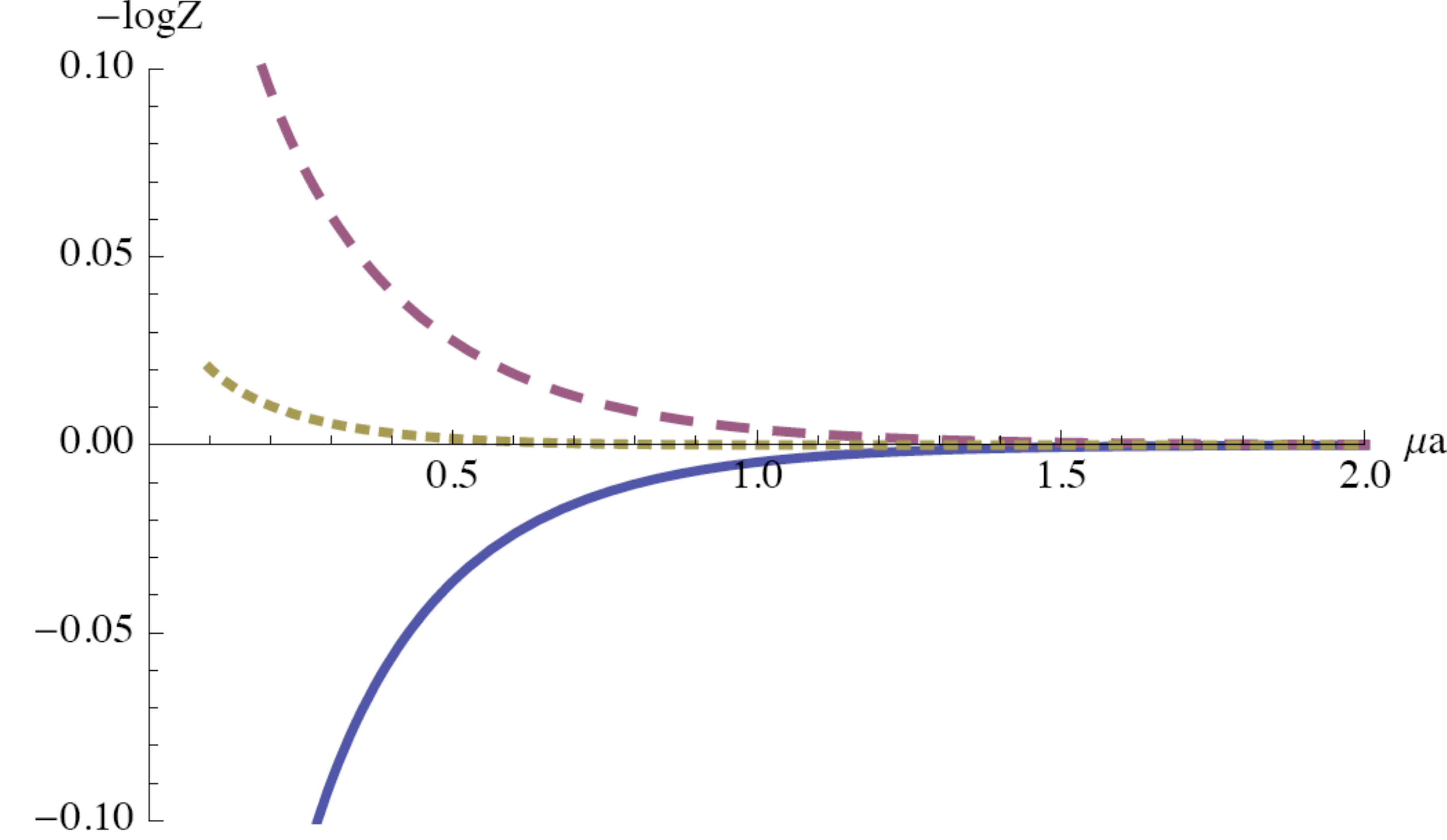}}
\end{center}
\caption{Free energy vs.\ $\mu a$ for a two dimensional slit.  Lower curve is $- \log Z^{(0)}$, upper dashed curve is $- \log Z^{(1)}$ and upper dotted curve is $-\log Z^{(2)}$.\label{2dslit}}
\end{figure}

The integrals involved in these formulae can be computed numerically
as a function of $\mu a$.
The direct term and the first two diffractive contributions are shown in
Fig.\ \ref{2dslit}.  Notice that the second order diffractive term is much smaller than 
the first order term, consistent with our expectation of the usefulness of 
the expansion (\ref{pertexp1}).

%%%%%%%%%%%%%%%%%%%%%%%%%%%%%%%%%%%%%%%%%%%%%%%%%%%%%%%%%%%%
\subsection{Slits in 4 dimensions\label{4dSlit}}
%%%%%%%%%%%%%%%%%%%%%%%%%%%%%%%%%%%%%%%%%%%%%%%%%%%%%%%%%%%%

We can now extend our results to the physical setting of four dimensions
by introducing two more
dimensions: a periodic Euclidean time dimension of size $\beta$ (representing the inverse temperature),
and a space dimension of size $L$ (representing the length measured 
along the edge of the slit).\footnote{While we use periodic boundary condition for the time direction, we will retain Dirichlet conditions for the spatial directions. In the limit of large $L$, we can replace summations over momenta along this spatial direction by integration. The distinction between Dirichlet conditions and periodic boundary conditions will not matter as $L \rightarrow \infty$.}
For large $\beta$ and $L$ the 4-dimensional partition function is an integral,
\be
- \log Z{}_{4d} = \beta L \int {d^2 \mu \over (2 \pi)^2} \, \left(- \log Z{}_{2d}\right)
\label{slit-diff14}
\ee
where we are interpreting the momentum in the extra dimensions as providing a Kaluza-Klein mass.
This means the energy per unit length for a slit in four dimensions is
\be
{E \over L} = \int_0^\infty {\mu d\mu \over 2 \pi} \, \left(- \log Z{}_{2d}\right)\,.
\label{slit-diff15}
\ee
The direct contribution is thus given, using (\ref{Z0renormalized}), by  
\be
{E^{(0)}}
=  - {\zeta(3) L \over 128 \pi a^2}
=  {L\over a^2} ~(-2.99 \times 10^{-3})\label{slit-diff16}
\ee
The diffractive contributions, obtained by integrating $-\log Z^{(n)}$ from
(\ref{slit-diff7}), (\ref{slit-diff11}), and (\ref{slit-diff12a}), (\ref{slit-diff12b}) are 
\be
{E^{(1)} } = {L\over a^2} ~(2.15 \times 10^{-3}), \hskip .2in{E^{(2)} } = {L\over a^2} ~(0.14 \times 10^{-3})
\label{slit-diff17}
\ee
The total value of the energy, to this order, is $-0.7\times 10^{-3} (L/a^2)$. The $1/a^2$ dependence of these results is, of course, fixed by dimensional analysis.

We may also note that the energy for a slit in arbitrary number of dimensions can be obtained by
extending the integration over $\mu$ in (\ref{slit-diff14}) to higher dimensions.

%%%%%%%%%%%%%%%%%%%%%%%%%%%%%%%%%%%%%%%%%%%%%%%%%%%%%%%%%%%%
\section{Perpendicular plates\label{perp}}
%%%%%%%%%%%%%%%%%%%%%%%%%%%%%%%%%%%%%%%%%%%%%%%%%%%%%%%%%%%%

\begin{figure}[!t]
\begin{minipage}{6.5cm}
\begin{center}
\includegraphics{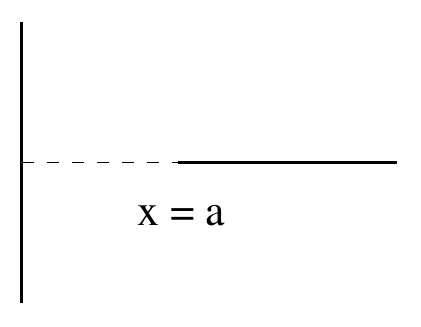}
\end{center}
\end{minipage}
\begin{minipage}{6cm}
\begin{center}
\includegraphics{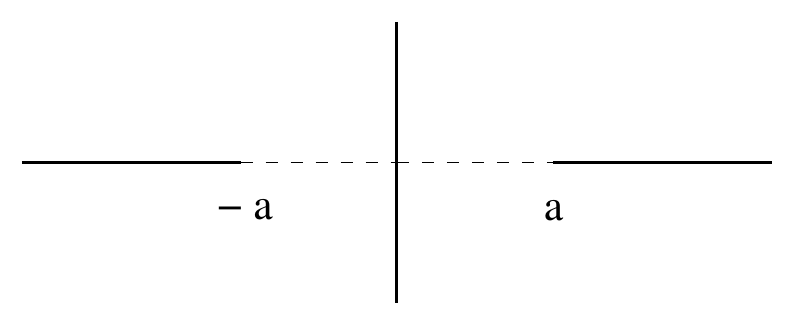}
\end{center}
\end{minipage}
\caption{On the left, perpendicular plates separated by a distance $a$.  On the right, the geometry described
by the odd modes in a slit.\label{PerpPlates}}
\end{figure}

In this section we study the ground state energy for two perpendicular plates
separated by a distance $a$.  The geometry is shown in the left panel of Fig.\ \ref{PerpPlates}.

Again our basic approach is to write a lower-dimensional effective
action in the gap between the plates, indicated by the dotted line in
the figure.  Fortunately this turns out to require very little effort.
The odd-parity modes in a slit that we studied in section \ref{slit} vanish at $x = 0$. 
Thus they are appropriate for the case of there being a plate perpendicular to the slit as
in the right panel of
Fig.\ \ref{PerpPlates}.  We are interested basically in one side of this geometry.
Notice however that the modes relevant for this case, namely, for
$0\leqslant x \leqslant a$ are in one-to-one correspondence
with the modes relevant for the slit $-a \leqslant x \leqslant a$.
The eigenvalues are also the same. 
The determinant is then given by the result for the odd-parity modes of the slit.
So to obtain the
Casimir of the two perpendicular plates, we can take the result for $-\log Z_{2d}$ discussed in the previous section,
but restricted to the odd-parity modes, and then integrate over $\mu$ for the appropriate number of transverse dimensions.\footnote{The result for the full geometry of the right panel
will require independent modes for the left and right sides of the vertical plate.
We must use modes $\sin (n\pi x/a)$, $\sin (m\pi x/a)$ with $m, n$ being independently chosen
integers. This will lead to a doubling of our results for that geometry.}
Carrying out the integrations in four dimensions, we find
\bea
E^{(0)}_\perp &=& {L\over a^2} ( - 11.96 \times 10^{-3})\nonumber\\
E^{(1)}_\perp &=& {L\over a^2} (~ 5.01  \times 10^{-3}), \hskip .35in
E^{(2)}_\perp = {L\over a^2} ( ~0.66  \times 10^{-3})\nonumber\\
E^{(3)}_\perp &=& {L\over a^2} ( ~0.16  \times 10^{-3}), \hskip .2in
E^{(4)}_\perp = {L\over a^2} ( ~0.05  \times 10^{-3})\label{slit-diff18}\\
E^{(5)}_\perp &=& {L\over a^2} ( ~0.01  \times 10^{-3})\nonumber
\eea
The total value for the $E_\perp$ up to this order is $-6.07(2)\times 10^{-3} (L/a^2)$. The terms in (\ref{slit-diff18}) have been evaluated using {\it Mathematica}. The $n$-th order term involves $2n+1$ integrals. As the number of integrals increases, the precision of the answers is lowered. We used several integration methods suitable for multi-dimensional integrals. Comparing results from different integration methods we estimate the error in our final answer for the total value to be  within $0.02 \times 10^{-3}$.

The Casimir energy for two perpendicular plates separated by a gap
has been numerically investigated by Gies and Klingm\"uller \cite{gies}.
Their calculation is done by considering 
a path integral representation for the propagator. When the two plates are present, all paths
which touch both plates must be considered as an overcounting of paths
and must be removed from the sum over paths.
This process, in a Monte Carlo evaluation of the path integral, then leads to corrections to the pure vacuum result and gives the Casimir energy.
Their final result for two perpendicular plates  is given as
\be
E_\perp = {L\over a^2} (-6.00 (2) \times 10^{-3})\label{slit-diff19}
\ee
Clearly our result is in very good agreement with the above value calculated in \cite{gies}.
%%%%%%%%%%%%%%%%%%%%%%%%%%%%%%%%%%%%%%%%%%%%%%%%%%%%%%%%%%%%
\section{Parallel plates\label{parallel}}
%%%%%%%%%%%%%%%%%%%%%%%%%%%%%%%%%%%%%%%%%%%%%%%%%%%%%%%%%%%%

\begin{figure}[!b]
\begin{center}
\includegraphics{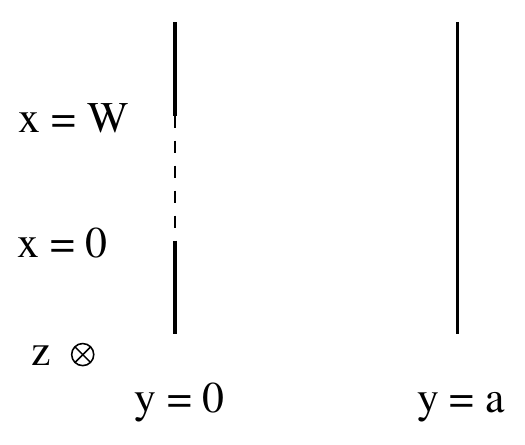}
\end{center}
\caption{Parallel plate geometry.  The $z$ axis is into the page.\label{ParallelFig}}
\end{figure}

Consider an infinite plate with a hole in it, parallel to a second
infinite plate with no hole.  Let the separation distance between the
plates be $a$.  The geometry is shown in Fig.\ \ref{ParallelFig}.

It is straightforward to study this situation along the lines of
section \ref{EffectiveAction}. 
We are interested in keeping $a$ finite but taking $b \rightarrow \infty$.
In this case
\be
M_L (x\vert x') + M_R (x\vert x') = 
  \sum_{\bf p}  \psi_{\bf m} ({\bf x}) \psi^*_{\bf m}({\bf x}')
~ p \left( 1 +\coth (ap)\right) 
\label{||platea}
\ee
Since $\psi_{\bf m}({\bf x})$ form a complete set of states, this is basically the operator
$ \sqrt{-\nabla^2} ( 1 + \coth a \sqrt{-\nabla^2})$. 
Once again, an alternative  way to arrive at the above equation is the following.
 A complete set of solutions to the bulk equations of
motion in the region between the plates is
\[
\phi_{\rm cl}({\bf x},y) = e^{i {\bf k} \cdot {\bf x}} \left(
A e^{-ky} + B e^{ky}\right)
\]
where the Dirichlet boundary conditions at $y = a$ fix
\[
B = - A e^{-2ka} \,
\]
Note that
\[
n \cdot \partial \phi_{\rm cl} \vert_{y = 0} = - \partial_y \phi_{\rm cl}
\vert_{y = 0} = k (A - B) e^{i {\bf k} \cdot {\bf x}} = k \, {A - B \over A + B}
\, \phi_{\rm cl} \vert_{y = 0}\,
\]
This means that we can identify
\[
n \cdot \partial \phi_{\rm cl} \vert_{y = 0} = {\sqrt{-\nabla^2} \over
\tanh(a\sqrt{-\nabla^2})} \, \phi_{\rm cl} \vert_{y = 0}
\]
The actions  (\ref{action1}), (\ref{action2}) are then given by
\bea
\label{HoleAction2}
&& S_L = \int d^{d-1}{\bf x} \, {1 \over 2} \phi_0 \sqrt{-\nabla^2} \phi_0 \\
\label{HoleAction3}
&& S_R = \int d^{d-1}{\bf x} \, {1 \over 2} \phi_0 {\sqrt{-\nabla^2}
\over \tanh(a\sqrt{-\nabla^2})} \phi_0
\eea
This can also be interpreted in the Hamiltonian language of section \ref{HamiltonianApproach}.
The wavefunctional on the left has the standard vacuum form
\[
\Psi_L[\phi_0] = \det{}^{-1/2}(-\Box_L) \, \exp\left( - \int d^{d-1}{\bf x} \, {1 \over 2} \phi_0 \sqrt{-\nabla^2} \phi_0\right)\,
\]
while the presence of the second plate modifies the wavefunctional on the right to
\[
\Psi_R[\phi_0] = \det{}^{-1/2}(-\Box_R) \, \exp \left( - \int d^{d-1}{\bf x} \, {1 \over 2} \phi_0 {\sqrt{-\nabla^2}
\coth(a \sqrt{-\nabla^2})} \phi_0\right)\,
\]
From either perspective, the partition function (\ref{Casimir}) is
\be
Z = \det{}^{-1/2}(-\Box_L) \det{}^{-1/2}(-\Box_R) \det{}^{-1/2} \left[ P \Big(
\sqrt{-\nabla^2} + {\sqrt{-\nabla^2} \coth (a\sqrt{-\nabla^2})}\Big) P \right]\, ,
\label{Z-||plate}
\ee
in agreement with (\ref{Casimir2}) and (\ref{||platea}).
Just to be clear: the first determinant is computed in the region $y < 0$ with a Dirichlet
boundary condition at $y = 0$.  The second determinant is computed in the region $0 < y < a$
with Dirichlet boundary conditions at $y = 0$ and $y = a$.  In the third determinant $P$ is a
projection operator onto the hole and $\sqrt{-\nabla^2}$ is the Laplacian on ${\mathbb R}^{d-1}$.

The first determinant, $\det{}^{-1/2}(-\Box_L)$, can be renormalized away.  The second determinant,
$\det{}^{-1/2}(-\Box_R)$, gives the Casimir energy that two parallel plates would have if there was
no hole.  In 3+1 dimensions this Casimir energy, per unit area, is given by the standard result
\be
\label{usual}
{E \over A} = - {\pi^2 \over 1440 a^3}\,
\ee
But in our case, for the sake of comparison with numerical results, it is important to specify
boundary conditions at infinity (meaning on the walls of the box shown in Fig.\ \ref{regions}).  In that
figure, if we increase the size of the hole until it reaches the walls of the box, the condition
that $\phi$ vanishes at the edge of the hole is carried over to a Dirichlet condition on the walls of the box.
So increasing the size of the hole is consistent provided we use Dirichlet boundary conditions at infinity.
That is, $\det{}^{-1/2}(-\Box_R)$ should be computed in a box of size $L_1 \times a \times L_3$,
where $L_1,\,L_3 \rightarrow \infty$ are the lengths in the $x,\,z$ directions.  This leads to subdominant terms in the Casimir
energy, namely
\be
E= - {\pi^2 \over 1440 a^3} L_1 L_3 + {\zeta (3) \over 64 \pi a^2} (2 L_1 + 2L_3) +
\cdots
\label{||plate1}
\ee
Other choices of boundary conditions at infinity are possible.  For instance, if we use periodic boundary conditions
in the $z$ direction, the term $\zeta (3)  L_1/ 32\pi a^2$ is absent and we'd have
\be
E= - {\pi^2 \over 1440 a^3} L_1 L_3 + {\zeta (3) \over 32 \pi a^2} L_3 + \cdots
\label{||plate1p}
\ee
For simplicity this is the case we will treat in the following.

Going back to (\ref{Z-||plate}), we note that all the dependence on the size and shape of the hole is captured by the third determinant, which we now
proceed to study.  To keep the discussion simple we take the hole to be a long slit of width $W$ and length $L = L_3$.
We expand the field in the slit in modes analogous to (\ref{OddModeFtns}), 
(\ref{EvenModeFtns}),
\be
\psi_{\omega k n} =  \sqrt{2\over W} ~\sin(n \pi x / W)
\label{||plate1a}
\ee
The slit is located at $0 \leq x \leq W$ so that $n = 1,2,3,\ldots$.

As in section \ref{PoleCut}, the operator we are interested in, namely,
\be
{\cal O} = P~ \sqrt{-\nabla^2} \Big( 1 +  \coth(a\sqrt{-\nabla^2})\Big) P
\label{||plate2}
\ee
can be decomposed into pole and cut contributions in this basis.  The free energy
$ F = {1 \over 2\beta} \, {\rm Tr} \, \log {\cal O}$
can then be expanded in powers of ${\cal O}^{\rm cut}$.  
The matrix elements of the projected operator in (\ref{||plate2}) are given by
\bea
{\cal O}_{mn} &=&  {4\over W} \int_0^W dx dx' ~ \int {dp\over 2\pi}
e^{ip(x-x')} \sin(n\pi x/W)
\sin(m\pi x'/W)~ f(p) \nonumber\\
f(p)&=& \sqrt{p^2 +\mu^2 } \left[ 1\over 1- \exp(-2a \sqrt{p^2 + \mu^2})\right]
\label{||plate4}
\eea
To evaluate the cut-terms arising from the square root factors, it is useful to write an integral representation for $f(p)$, namely,
\be
f(p) = {1\over \pi} \int_{-\infty}^\infty d\alpha {p^2 + \mu^2 \over
(\alpha^2 +p^2 + \mu^2)}~\left[ {1\over 1- \exp(2a \alpha i -\epsilon )}
\right]
\label{||plate5}
\ee
The exponent $\epsilon$ pushes the poles at 
$\alpha = n \pi /a$ to the lower half-plane.
We can evaluate the $\alpha$-integral by completing the contour in the upper half-plane; only the pole at $\alpha = i \sqrt{p^2 + \mu^2}$ contributes and
the equivalence with (\ref{||plate4}) can be easily verified.
Carrying out the integration over $x, x'$, we then find
\bea
{\cal O}_{mn} &=& {4\over W} \int {dp\over 2\pi} f(p) H(q, q')\nonumber\\
H(q, q')&=& {1\over 4}  \Biggl[ {1+e^{i(q-q')W}- e^{i(p-q)W} - e^{-i(p-q')W}
\over (p-q+i\epsilon) (p-q'-i\epsilon)}\nonumber\\
&&\hskip .2in
- (q\rightarrow -q, q'\rightarrow q')
-(q\rightarrow q, q'\rightarrow -q') + (q\rightarrow -q, q'\rightarrow -q')
\Biggr]
\label{||plate6}
\eea
where $q= n\pi/W, q'= m\pi/W$.
The integration over $p$ can now be done.
For a term with $1+e^{i(q-q')W}- e^{i(p-q)W}$, we nee to close the contour in the upper half-plane, for a term with $e^{-i(p-q')W}$, we need to close in the lower half-plane.
There will be pole contributions from the denominators $p-q +i\epsilon$ and
$p-q'-i\epsilon$. These are identical to what we named the pole terms
in ${\cal O}_{mn}$. There will also be terms from the poles of $f(p)$.
The latter will correspond to the cut-terms we are seeking.
The evaluation of the $p$-integral then leads to
${\cal O}_{mn} = {\cal O}^{\rm pole}_{mn} + {\cal O}^{\rm cut}_{mn}$, with\footnote{Previous versions of this paper had an overall sign error in ${\cal O}^{\rm cut}_{mn}$.}
\bea
{\cal O}^{\rm pole}_{mn} &=& { 2 \omega (q) \over 1 - e^{-2 a\, \omega (q)} }~\delta_{mn} = 2 f(q)~ \delta_{mn}\nonumber\\
{\cal O}^{\rm cut}_{mn}&=& {4  q q'\over \pi W} \, [ 1 + (-1)^{m+n}]\,  \Delta ( a, q, q')\label{||plate6a}\\
\Delta (a, q, q') &=& \int_0^\infty d \lambda \left[ { f (\lambda )\over (q^2 -\lambda^2) (q'^2 - \lambda^2)}
+{ f (q)\over (q'^2 - q^2) (\lambda^2 - q ^2)} +
{ f (q' )\over (q^2 - q'^2) (\lambda^2 - q'^2)}\right]
\nonumber
\eea
where $\omega (q) = \sqrt {q^2 +\mu^2}$.
%%%%%%%%%%%%%%%%%%%%%%%%%%%%%%%%%%%%%%%%%%%%%%%%%%%%%%%%%%%%%%%%%%%%%%
\subsection{The direct contribution (lowest order)}
%%%%%%%%%%%%%%%%%%%%%%%%%%%%%%%%%%%%%%%%%%%%%%%%%%%%%%%%%%%%%%%%%%%%%%
The direct contribution to the free energy is
given by
\bea
\beta F& = & {1 \over 2} \, {\rm Tr} \, \log {\cal O}^{\rm pole} \\
\label{LowestParallel}
& = & {1 \over 2} \, {\rm Tr} \, \log (2 \sqrt{-\nabla^2_D}) - {1 \over 2} \, {\rm Tr} \, \log
\left(1 - e^{-2 a \sqrt{-\nabla^2_D}}\right)
\eea
where $\nabla^2_D$ is the Laplacian in the slit with Dirichlet boundary conditions at $x=0$ and $x = W$.
The first term has UV divergences but is independent of $a$.  In fact it is just the lowest order energy in a
single slit which
we studied in section \ref{4dSlit}.  There we found that the energy per unit length for a slit of
width $W$ is\footnote{One
can obtain this directly, as ${E / L} = (1 /\beta L) \, {1 \over 4} \, {\rm Tr} \, \log (-\nabla^2_D)$,
using the results in appendix \ref{HeatKernels}.}
\be
{E \over L} = - {\zeta(3) \over 32 \pi W^2} \label{||plate7}
\ee
The dependence on the separation between plates is captured by the second term in (\ref{LowestParallel})
which is finite in the UV.  Including both terms, we have the finite (renormalized) energy per unit length
\be
{E \over L} = - {\zeta(3) \over 32 \pi W^2} - {1 \over 2} \int {d^2 k \over (2\pi)^2} \sum_{n = 1}^\infty \,
\log \left(1 - e^{-2a\sqrt{k^2 + (n\pi/W)^2}}\right)\label{||plate8}
\ee

This expression can be studied in various limits.  As $W \rightarrow 0$ the second term makes an exponentially
small correction, and we have
\be
{E \over L} \approx - {\zeta(3) \over 32 \pi W^2} + {1 \over 8 a W} e^{- 2 \pi a / W}
\label{||plate9}
\ee
On the other hand as $W \rightarrow \infty$ the second term dominates.  To study it in this limit we use
the Euler-Maclaurin summation formula,
\be
\sum_{n=1}^\infty f\Big({n \over W}\Big) = W \int_0^\infty dx \, f(x) - {1 \over 2} f(0) - \sum_{m=1}^\infty
{B_{2m} \over (2m)!} {f^{(2m-1)}(0) \over W^{2m-1}} \label{||plate10}
\ee
This leads to
\be
{E\over L } = {\pi^2 W  \over 1440 a^3} - {\zeta(3)  \over 32 \pi a^2} + {\cal O}(1/W)
\label{||plate11}
\ee
Adding this result to the bulk contribution (\ref{||plate1p}), with $L_1$ and $L_3 = L$ taken to be large,
we find, for the lowest order or direct contribution to the energy,
\be
{E^{(0)} \over L} = - {\pi^2 (L_1-W)  \over 1440 a^3} +\cdots
\label{||plate3}
\ee
The terms proportional to $1/a^2$ cancel out.\footnote{This cancellation depends on the boundary conditions
at infinity used in (\ref{||plate1p}), namely Dirichlet in $x$ and periodic in $z$.  With a different choice of
boundary conditions at infinity the $1/a^2$ terms would not cancel.}
Also the usual Casimir energy per unit area (\ref{usual}) in the region corresponding
to the slit is canceled out and only the 
facing area of the two plates $L (L_1 -W)$ appears in $E^{(0)}$.
%%%%%%%%%%%%%%%%%%%%%%%%%%%%%%%%%%%%%%%%%%%%%%%%%%%%%%%%%%%%%%%%%%%%%%
\subsection{First diffractive contribution (first order)}
%%%%%%%%%%%%%%%%%%%%%%%%%%%%%%%%%%%%%%%%%%%%%%%%%%%%%%%%%%%%%%%%%%%%%%
The diffractive contribution to the 2d free energy arises from the expansion
\be
- [ \log Z  - \log Z^{(0)}]  = {1\over 2} \Tr \log \left( \delta_{mn} + {2 q q' \over \pi W}
[ 1 + (-1)^{m+n}] \,{1\over f(q)}~ \Delta (a, q, q')\right)\label{||plate12}
\ee
We can easily work out the higher order terms from this. In the case when $W$ is large,
\bea
-\log Z^{(1)} &=& {2 L \over \pi^2} \int_0^\infty dq\, q^2\, {1- e^{- 2 a \omega(q)} \over \omega(q)}
\, \Delta (a, q, q)\nonumber\\
\Delta (a, q, q) &=& {1\over 2 q} {\del \over \del q} \int_0^\infty d \lambda\,
{f(\lambda ) - f (q) \over \lambda^2 - q^2}\label{||plate13}
\eea
The $n$-th order term is given by
\bea
-\log Z^{(n)} &=& (-1)^{n+1} {2^n L \over n \pi^{2n} }\int_0^\infty \prod^n_i dq_i \, q_i^2~
~{\Delta (a)\over f} \ast {\Delta (a)\over f} \ast \cdots \ast {\Delta (a)\over f}\nonumber\\
{\Delta (a)\over f} \ast  \cdots \ast {\Delta (a)\over f}
&=& {\Delta (a, q_1, q_2)\over f(q_1)}\, { \Delta (a, q_2, q_3)\over f(q_2)} \cdots {\Delta (a, q_n, q_1)
\over f(q_n)}
\label{||plate14}
\eea
As in the case of the slit and the perpendicular plates, the renormalized expressions are obtained by subtracting the $a\rightarrow \infty$ limit.
\be
-\log Z^{(n)}_{\rm ren} = (-1)^{n+1} {2^n L \over n \pi^{2n} }\int_0^\infty \prod^n_i dq_i \, q_i^2~
\left[ {\Delta (a)\over f} \ast {\Delta (a)\over f} \ast \cdots \ast {\Delta (a)\over f}
- (a \rightarrow \infty )\right]
\label{||plate15}
\ee
The energy for the case of four dimensions can now be obtained by integration over
$\mu$,
\be
E = \int {\mu d\mu \over 2\pi} \Bigl( - \log Z_{\rm ren}\Bigr)
\label{||plate16}
\ee
Evaluating the integrals numerically, we find, for the first few orders,
\bea
E^{(1)} &=& {L \over a^2} ~(~5.54 \times 10^{-3}), \hskip .2in
E^{(2)} =  {L \over a^2} ~(~0.80\times 10^{-3}),\nonumber\\
E^{(3)} &=&  {L \over a^2} ~(~0.19 \times 10^{-3}), \hskip .27in
E^{(4)} =  {L \over a^2} ~(~0.05  \times 10^{-3}),\nonumber\\
E^{(5)} &=&  {L \over a^2} ~(~0.01 \times 10^{-3})\label{||plate17}
\eea
The value for the diffractive contribution to the energy, up to this order, is $E^{\rm diffr} = 6.59 \times 10^{-3} (L/a^2)$.  To obtain the total energy associated with the edges of the slit
we combine the direct contribution which appears in (\ref{||plate11}) with the diffractive contribution computed here, to find\footnote{Since we are interested in the energy associated
with the edges of the slit, we leave out the usual Casimir energy (\ref{||plate3}) associated with the facing area of the two plates.  We also leave out the $\zeta(3) L_3 / 32 \pi a^2$ term in (\ref{||plate1p})
since, rather than being associated with the slit, it is associated with the corners which appear at infinity when a Dirichlet condition at infinity in the $x$ direction is imposed.}
\be
E^{\rm slit} = - {\zeta(3) L \over 32 \pi a^2} + E^{\rm diffr} \sim -5.37 \times 10^{-3} \, {L \over a^2}
\ee
This result is for a slit of finite, although large, width. There is no direct
comparison to other methods of calculation available. 
However, the case of two parallel plates, one of which is semi-infinite, provides a point of comparison.
The Casimir energy for this geometry has been numerically investigated by Gies and Klingm\"uller \cite{gies}, by the world-line method of
subtracting out the paths which touch both plates.
Their final result for two parallel plates, one of which is semi-infinite, is given as
\be
E^{\rm edge} = - {\gamma \over 2} {L \over a^2}
+\cdots
\label{||plate18}
\ee
where $\gamma = 0.00523(2)$.

We have a slit of finite, although large, width.
Thus there are two edges to the slit, each of length $L$ which must be considered.
If one were to remove paths from a sum-over-paths formula for the propagator,
all paths which touch both edges must be removed.
The calculation in \cite{gies} is for a semi-infinite plate to begin with and hence
paths which touch on the edge which is far away from  $x=0$ are not removed.
Thus our result must be divided by $2$ for the edge terms to get a proper comparison, and we find
\be
E^{\rm edge} = - {\gamma \over 2} {L \over a^2} \quad {\rm with} \quad \gamma \sim 5.37 \times 10^{-3}
\ee
This value for $\gamma$ is in good agreement with the result of Gies and Klingm\"uller \cite{gies}.  The inclusion of higher order diffractive terms would further decrease
our value for $\gamma$ and presumably bring it closer to the world-line result of \cite{gies}.

An exact calculation of the Casimir energy of a parabolic cylinder next to an infinite plate has been done by
Graham {\it et al} \cite{MIT2}. A particular limit of this gives the result for the case we are studying, namely, a semi-infinite plate next to
a parallel plate.
The result in \cite{MIT2} is ${1\over 2}\gamma = 0.0025$, again in agreement with our result
and with \cite{gies}.

\section{Summary}

We have developed a method for calculating Casimir energies, including diffractive contributions which can arise
from apertures on plates and other boundary elements of the geometry. This involves the functional integration over a lower dimensional field theory defined on such apertures. The relevant kinetic operator has an interesting structure. In the simplest cases it is of the form $\sqrt{-\nabla^2}$, similar to what occurs in the wave functional of the field, but there are important modifications based on the geometry of the situation.
In all cases, the operator acts on functions which have support only on the apertures.
The matrix elements of the operator allow a clean separation of diffractive contributions from direct (or ray optics) contributions. To evaluate the relevant functional integrals we expanded in powers of the diffractive contribution.
This seems to be a good approximation even though there is no explicit small parameter in the
problem. 

In this paper we focused on the Casimir energy for some special cases: a single slit, two parallel plates,
one of which has a long slit in it, and two perpendicular plates separated by a gap.
In the latter two cases numerical calculations based on world line methods have been performed.  Our results
can be compared, and in both cases the agreement is quite good.  But the method we have developed
is quite general and can be applied to a variety of different geometries.  For instance
it can be easily generalized to arbitrary dimensions.  In fact working in $d$ dimensions might justify the perturbation series, as an expansion in powers of $1/d$.  The method could also be extended to include finite temperature effects, which would allow a comparison with the results of \cite{Gies2}. 

\bigskip
\centerline{\bf Acknowledgements}
We thank Rashmi Ray for collaboration at an early stage of this work.
We are also grateful to Janna Levin and Alexios Polychronakos for valuable discussions.  This work was supported by U.S.\ National Science
Foundation grants PHY-0855582, PHY-0758008 and PHY-0855515
and by PSC-CUNY awards.  DK is grateful to the Aspen Center for Physics where part of this work was completed.

\appendix
\section{Exact mode functions on an interval and half-line}\label{appendix1}

In the formalism we have developed, effects associated with a hole are
captured by non-local differential operators of the form
$P F P$
where $P$ is a projection operator onto the hole and $F$ is some
function of the Laplacian on ${\mathbb R}^{d-1}$.  Diagonalizing such
operators is in general difficult and we were forced to resort to
perturbation theory.  However one can find the exact eigenfunctions
numerically.  Also in some cases an analytical treatment is
possible, using results obtained long ago by Malyuzhinets for
scattering from a wedge \cite{Malyuzhinets,MalyuzhinetsReview}.  Here we collect some of these
results.  Besides illustrating the non-perturbative features
of the problem, our motivation is to provide evidence that a
perturbative treatment should be reliable.

\begin{figure}[!b]
\begin{center}
\includegraphics[width=12cm,height=6cm]{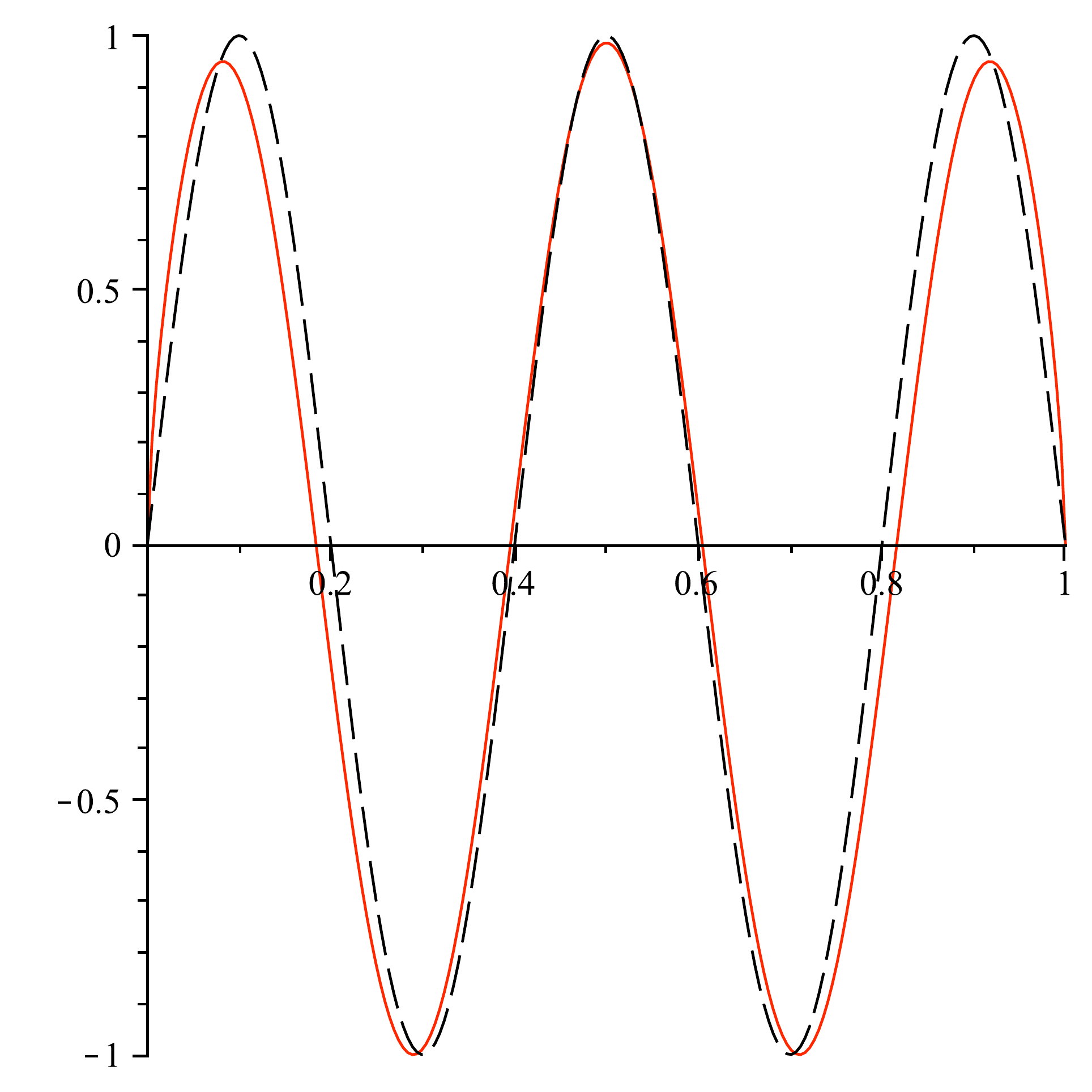}
\end{center}
\caption{The 5${}^{\it th}$ mode in a slit.  The red solid curve is the exact eigenfunction.  The black dashed
curve is $\sin (5 \pi x)$.\label{NumericalMode}}
\end{figure}

First consider the operator $PFP$ where $F = \sqrt{-d^2/dx^2}$ and $P$
is a projection operator on the unit interval $[0,1]$.  One can study
this numerically, starting from a finite difference approximation to the
Laplacian in position space.
\be
\label{FiniteDiff}
(F^2)_{ii} = 2 \qquad (F^2)_{i,i+1} = (F^2)_{i,i-1} = -1
\ee
In this basis
\[
P = \left(
\begin{array}{c|c|c}
0 & 0 & 0 \\
\hline
0 & \identity & 0 \\
\hline
0 & 0 & 0
\end{array}
\right)
\]
One can take the square root of (\ref{FiniteDiff}) numerically and diagonalize $PFP$.
A typical eigenfunction is shown in Fig.~\ref{NumericalMode}.
The exact eigenfunction clearly vanishes at the edges of the interval and approaches a plane
wave in the middle.  It is well-approximated by the Dirichlet modes $\sin(n \pi x)$ that we
used as the basis for our perturbation series.  Indeed the only significant difference between the
exact and perturbative modes is that the exact modes go to zero more steeply at the edges
of the interval.

To study this edge behavior we consider
$F = \sqrt{-{d^2 \over dx^2} + \mu^2}$ and take $P$ to be a projection operator onto a half-line.
\be
\label{HalfLine}
P \, f(x) = \left\lbrace
\begin{array}{ll}
f(x) & x > 0 \\
0 & \hbox{\rm otherwise}
\end{array}
\right.
\ee
This describes a semi-infinite ``plate'' embedded in two Euclidean
spacetime dimensions.  As this geometry has no adjustable parameters,
up to renormalization the energy of such a plate vanishes,
and in this sense there is nothing interesting to calculate.  Instead our
motivation for studying this geometry is that we will be able to diagonalize
$PFP$ analytically and show that the exact eigenfunctions have $\sqrt{x}$ behavior
as $x \rightarrow 0$.  We will show this in two different ways: first using
Laplace transforms, then by solving the wave equation following Malyuzhinets.

%%%%%%%%%%%%%%%%%%%%%%%%%%%%%%%%%%%%%%%%%%%%%%%%%%%%%
\subsection{Laplace transform}
%%%%%%%%%%%%%%%%%%%%%%%%%%%%%%%%%%%%%%%%%%%%%%%%%%%%%

The eigenvalue problem we wish to solve is
\be
\label{EvalProb}
P F P \phi(x) = \lambda \phi(x)\,.
\ee
One way to solve (\ref{EvalProb}) is to find a function $\phi(x)$ such that
\begin{enumerate}
\item
$\phi(x)$ has support for $x > 0$, so that $P \phi = \phi$
\item
$(F - \lambda) \phi$ has support for $x < 0$, so that $P (F - \lambda) \phi = 0$
\end{enumerate}
Suppose we represent $\phi(x)$ using an inverse Laplace transform.
\be
\label{Laplace}
\phi(x) = \int_{-i\infty}^{i\infty} {ds \over 2 \pi i} \, e^{sx} \tilde{\phi}(s)
\ee
If $\tilde{\phi}$ is analytic for ${\rm Re} \, s > 0$, then $\phi(x)$ will vanish for $x < 0$.
Likewise if $\tilde{\phi}$ is analytic for ${\rm Re} \, s < 0$, then $\phi(x)$ will vanish for $x > 0$.
So we need to find a function $\tilde{\phi}(s)$ such that
\begin{enumerate}
\item
$\tilde{\phi}(s)$ is analytic for ${\rm Re} \, s > 0$
\item
$(\sqrt{\mu^2 - s^2} - \lambda) \tilde{\phi}(s)$ is analytic for ${\rm Re} \, s < 0$
\end{enumerate}
It's convenient to work on the covering space of the cut $s$ plane, making a change of variables
$s = i \mu \sinh z$.  We also set $\lambda = \mu \cosh \xi$.\footnote{The $\delta$-function normalizeable spectrum
of $PFP$ is $\lambda \in [\mu ,\infty)$ corresponding to $0 \leq \xi < \infty$.}  Then (\ref{Laplace}) becomes
\[
\phi(x) = {\mu \over 2 \pi} \int_{-\infty}^\infty dz \, \cosh z \, e^{i \mu x \sinh z} \tilde{\phi}(z)\,.
\]
If $\tilde{\phi}(z)$ is analytic in the strip $- \pi \leq {\rm Im} z \leq 0$ and
satisfies
\[
\tilde{\phi}(z) = \tilde{\phi}(-z-i\pi)
\]
then we can replace
\[
\int_{-\infty}^\infty dz \rightarrow {1 \over 2} \left(\int_{-\infty}^\infty + \int_\infty^{\infty - i\pi}
+ \int_{\infty - i\pi}^{-\infty - i \pi} + \int_{-\infty - i \pi}^{-\infty}\right) dz
\]
and $\phi(x)$ will vanish for $x < 0$.  (The conditions on $\phi(z)$ correspond to the requirement that
$\tilde{\phi}(s)$ is analytic and single-valued for ${\rm Re} \, s \geq 0$.)  Likewise if $\tilde{\phi}(z)$
is analytic in the strip $0 \leq {\rm Im} \, z \leq \pi$ and satisfies
\[
\tilde{\phi}(z) = \tilde{\phi}(-z + i \pi)
\]
(corresponding to the requirement that $\tilde{\phi}(s)$ is analytic
and single-valued for ${\rm Re} \, s < 0$) then $\phi(x)$ will vanish
for $x > 0$.  So corresponding to the conditions on $\tilde{\phi}(s)$,
we need a function $\tilde{\phi}(z)$ such that
\begin{enumerate}
\item
$\tilde{\phi}(z)$ is analytic for $- \pi \leq {\rm Im} \, z \leq 0$ and satisfies
\be
\label{Difference1}
\tilde{\phi}(z) = \tilde{\phi}(-z-i\pi)
\ee
\item
$(\cosh z - \cosh \xi) \tilde{\phi}(z)$ is analytic for $0 \leq {\rm Im} \, z \leq \pi$ and satisfies
\be
\label{Difference2}
(\cosh z - \cosh \xi) \tilde{\phi}(z) = - (\cosh z + \cosh \xi) \tilde{\phi}(-z+i\pi)
\ee
\end{enumerate}
A solution to this system of equations was obtained by Malyuzhinets \cite{Malyuzhinets,MalyuzhinetsReview}.
Define
\[
\Psi_\pi(z) = \psi_\pi\big(-iz + {3 \pi \over 2} + i \xi\big) \psi_\pi\big(-iz + {3 \pi \over 2} - i \xi\big)
              \psi_\pi\big(-iz - {\pi \over 2} + i \xi\big) \psi_\pi\big(-iz - {\pi \over 2} - i \xi\big)
\]
where
\[
\psi_\alpha(z) = \exp - {1 \over 2} \int_0^\infty dt \, {\cosh(tz) - 1 \over t \cosh (t \pi/2) \sinh(2 \alpha t)}\,.
\]
Then the solution is
\be
\label{LaplaceSolution}
\tilde{\phi}(z) = {1 \over (\sinh z - i \epsilon)^2 - \sinh^2 \xi} \, \Psi_\pi(z)\,.
\ee
To see this, note that
$\Psi_\pi(z)$ was constructed to satisfy (\ref{Difference1}), (\ref{Difference2}) all by itself.  Moreover the prefactor
$1/(\sinh^2z - \sinh^2\xi)$ is invariant under $z \rightarrow -z \pm i \pi$, so $\tilde{\phi}(z)$
also satisfies (\ref{Difference1}), (\ref{Difference2}).  It only remains to check the analyticity conditions.
In the strip $-\pi \leq {\rm Im} \, z \leq \pi$ one can show that $\Psi_\pi(z)$ has no poles.  It does, however, have zeroes at
$z = \pm \xi + i \pi$.  In the same strip the prefactor has poles at $z = \pm \xi + i \epsilon$ and
$z = \pm \xi + i \pi - i \epsilon$, but the latter poles cancel against the zeroes of $\Psi_\pi$.  So in the
strip $\tilde{\phi}$ has poles at $z = \pm \xi + i \epsilon$.
That is, $\tilde{\phi}(z)$ is analytic for $- \pi \leq {\rm Im} \, z \leq 0$ and, when multiplied by $\cosh z - \cosh \xi$, it
becomes analytic for $0 \leq {\rm Im} \, z \leq \pi$.

One comment on this solution is in order.  Up to a normalization the prefactor in (\ref{LaplaceSolution})
is the Laplace transform of $\theta(x) \sin kx$, where $k = M \sinh \xi$ -- exactly the
modes with Dirichlet boundary conditions that were the starting point for our perturbation theory.  So in the
case of a semi-infinite plate, diffractive corrections to the perturbative modes are given by the Malyuzhinets
function $\Psi_\pi(z)$.

%%%%%%%%%%%%%%%%%%%%%%%%%%%%%%%%%%%%%%%%%%%%%%%%%%%%%
\subsection{Wave equation in a wedge}
%%%%%%%%%%%%%%%%%%%%%%%%%%%%%%%%%%%%%%%%%%%%%%%%%%%%%

Another approach to diagonalizing $PFP$ more closely makes contact with the
original work of Malyuzhinets.  Consider a semi-infinite plate in two Euclidean dimensions, and let's study the wavefunctional for the field on a `hole'
which is a half-line making an angle $\alpha$ with respect to the plate.
\begin{figure}
\begin{center}
\includegraphics{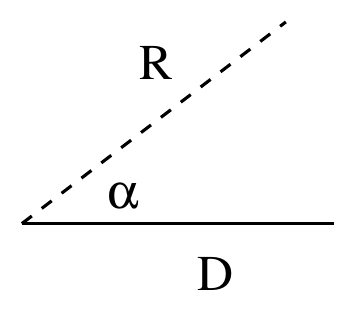}
\end{center}
\caption{The wedge geometry.  We impose Dirichlet boundary conditions on the plate at $\theta = 0$.  To diagonalize
the operator $PFP$ we impose Robin boundary conditions on the hole at $\theta = \alpha$.\label{wedge}}
\end{figure}
In general the surface actions appearing in (\ref{action1}), (\ref{action2}) can be written as
\[
S = \int_{\rm hole} {1 \over 2} \phi_{\rm cl} \, n \cdot \partial \, \phi_{\rm cl}
\]
where $\phi_{\rm cl}$ is a solution to $(-\Box + \mu^2) \phi_{\rm cl} = 0$ with the boundary conditions
$\phi_{\rm cl} = 0$ at $\theta = 0$, $\phi_{\rm cl} = \phi_0$ at $\theta = \alpha$.
Rather than specify the value of $\phi_0$, suppose we impose $n \cdot \partial \phi_{\rm cl} = \lambda \phi_{\rm cl}$ at $\theta = \alpha$.
That is, suppose we solve the system
\bea
\nonumber
(-\Box + \mu^2)\phi_{\rm cl} &=& 0 \\
\label{MalBC}
\phi_{\rm cl} &=& 0 \hskip .4in {\rm at}~ \theta = 0 \hskip .1in {\rm (Dirichlet ~boundary ~condition)} \\
\nonumber
n \cdot \partial \phi_{\rm cl} &=& \lambda \phi_{\rm cl} \hskip .2in {\rm at}~ \theta = \alpha 
\hskip .1in
{\rm (Robin ~boundary ~condition)}
\eea
To make contact with the problem of diagonalizing $PFP$, note that when $\alpha = \pi$ we have another expression for the surface
action, namely
\[
\int_{\rm hole} {1 \over 2} \phi_0 \, PFP \, \phi_0\,.
\]
So when $\alpha = \pi$, the value of the field along the Robin boundary $\phi_{\rm cl}(r,\theta = \alpha)$ is a solution to
$PFP \, \phi = \lambda \phi$.

Fortunately (\ref{MalBC}) is exactly the problem studied by Malyuzhinets \cite{Malyuzhinets,MalyuzhinetsReview}.
For general $\alpha$, and denoting $\lambda = \mu \cosh \xi$, the solution is
\[
\phi_{\rm cl}(r,\theta) = \int_{\gamma_+ \, \cup \, \gamma_-} {dz \over 2 \pi i} \, e^{-m r \cos(z - \theta)} f(z) g(z)
\]
where
\[
f(z) = \psi_\alpha(z + \alpha + i \xi) \psi_\alpha(z + \alpha - i \xi) \psi_\alpha(z - \alpha + i \xi)
\psi_\alpha(z - \alpha - i \xi)
\]
is the function introduced by Malyuzhinets and
\[
g(z) = {1 \over \sin^2(\nu z) - \sin^2(\nu \beta)}
\]
is chosen to obtain the correct asymptotic behavior at large $r$.  Here $\nu = \pi/2\alpha$ and
$\beta = {\pi \over 2} - \alpha + i \xi$.  The contour $\gamma_+$ starts at $z = 2 \pi + i \infty$, descends towards
the real axis, moves to $2 \pi$ to the left while staying above any singularities of the integrand, and returns to $+i \infty$,
while $\gamma_-$ is the mirror image of $\gamma_+$ under $z \rightarrow -z$.

One can extract the asymptotic behavior of the solution from the contour integral representation
\cite{Malyuzhinets,MalyuzhinetsReview}.  Along the Robin boundary $\phi_{\rm cl}$ has
plane-wave behavior at large $r$, while near the origin it has power-law behavior $\phi_{\rm cl}(r) \sim r^{\pi / 2 \alpha}$.
Setting $\alpha = \pi$, this means eigenfunctions of $PFP$ have plane-wave behavior far from the edge, while
near the edge they vanish like $\sqrt{r}$ as $r \rightarrow 0$.

%%%%%%%%%%%%%%%%%%%%%%%%%%%%%%%%%%%%%%%%%%%%%%%%%%%%%%%%%%%
\section{Heat kernels\label{HeatKernels}}
%%%%%%%%%%%%%%%%%%%%%%%%%%%%%%%%%%%%%%%%%%%%%%%%%%%%%%%%%%%

For a free scalar field of mass $m$ in Euclidean space, with some
number of periodic dimensions and some number of Dirichlet directions,
the partition function is
\[
Z = e^{-\beta F} = \det{}^{-1/2}(-\Box + m^2)\,.
\]
Here $\beta$ is just the periodicity around some ``Euclidean time'' direction.
We can represent
\bea
\nonumber
\beta F & = & {1 \over 2} {\rm Tr} \, \log (-\Box + m^2) \\
\label{IntegralRep}
& = & - {1 \over 2} \int_{\epsilon^2}^\infty {ds \over s} \, {\rm Tr} \, e^{-s(-\Box + m^2)}
\eea
where $\epsilon \rightarrow 0$ serves as a UV regulator.  To compute
the (trace of the) heat kernel
\[
K(s) = {\rm Tr} \, e^{s \, \Box}
\]
we use the fact that $\Box = \sum_i {\partial^2 \over \partial x_i^2}$
where the eigenvalues of $\partial_i^2$ are
\beas
&&\hbox{\rm $x_i$ periodic with period $L_i$} \quad \Rightarrow \quad
\hbox{\rm eigenvalues $-\left({2\pi n \over L}\right)^2$, $n \in {\mathbb Z}$} \\
&&\hbox{\rm $x_i$ Dirichlet with size $L_i$} \quad \Rightarrow \quad
\hbox{\rm eigenvalues $-\left({\pi n \over L}\right)^2$, $n \in {\mathbb N}$}
\eeas
This means that the heat kernel factorizes, $K(s) = \prod_i K_i$, where
\beas
&&\hbox{\rm $x_i$ periodic} \quad \Rightarrow \quad
K_i = K_P(s,L_i) = \sum_{n \in {\mathbb Z}} e^{-s(2\pi n / L_i)^2} \\
&&\hbox{\rm $x_i$ Dirichlet} \quad \Rightarrow \quad
K_i = K_D(s,L_i) = \sum_{n \in {\mathbb N}} e^{-s(\pi n / L_i)^2}
\eeas
Note that
\be
\label{Dirichlet}
K_D(s,L) = {1 \over 2} \left(K_P(s,2L) - 1\right)
\ee
By Poisson resummation
\be
\label{Poisson}
K_P(s,L) = {L \over \sqrt{4\pi s}} \sum_{n \in {\mathbb Z}} e^{-L^2 n^2 / 4s}
= \theta_3\left(0,{i 4 \pi s \over L^2}\right)
\ee
The expressions (\ref{Dirichlet}), (\ref{Poisson}) make it clear that as $s \rightarrow 0$ we have
\[
K(s) \sim K_0(s) = \prod_P {L_i \over \sqrt{4\pi s}} \,
\prod_D \left({L_i \over \sqrt{4 \pi s}} - {1 \over 2}\right)\,.
\]
This isolates the UV divergence: when we use this small-$s$ behavior
in the integral (\ref{IntegralRep}) we get a divergence as
$\epsilon \rightarrow 0$.  To get a finite answer we just subtract off
the contribution of $K_0$.  This has the interpretation of
renormalizing the various bulk and boundary cosmological constants.
(Terms in $K_0$ are proportional to the total volume, or the volumes
of various walls or corners.)  After making the subtraction we can set
$\epsilon = 0$.  So the renormalized answer is
\be
\label{final}
\beta F = - {1 \over 2} \int_0^\infty {ds \over s} \, e^{-s m^2} \left[\prod_P K_P(s,L_i) \,
\prod_D K_D(s,L_i) - \prod_P {L_i \over \sqrt{4\pi s}} \,
\prod_D \left({L_i \over \sqrt{4 \pi s}} - {1 \over 2}\right)\right]
\ee
where just for completeness
\beas
K_P(s,L) & = & {L \over \sqrt{4 \pi s}} + {L \over \sqrt{\pi s}} \sum_{n = 1}^\infty e^{-L^2 n^2 / 4s} \\
K_D(s,L) & = & {L \over \sqrt{4 \pi s}} - {1 \over 2} + {L \over \sqrt{\pi s}} \sum_{n = 1}^\infty e^{-L^2 n^2 / s}\,.
\eeas

\end{document}